\documentclass[%
 twocolumn,prb,
 superscriptaddress,
 amsmath,amssymb,
 aps,
floatfix,
]{revtex4-2}

\usepackage{graphicx}
\usepackage{dcolumn}
\usepackage{bm}
\usepackage{grffile} 
\usepackage{subfigure}
\usepackage[marginal]{footmisc}
\usepackage{epstopdf}
\usepackage{amssymb}
\usepackage{mathrsfs}
\usepackage{amsmath,mathtools,amsfonts,bm,graphicx}
\usepackage{upgreek}
\usepackage{color}
\usepackage{etoolbox}
\usepackage[colorlinks=true,linkcolor=blue,urlcolor=blue,citecolor=blue]{hyperref} 
\usepackage{multirow}
\usepackage[normalem]{ulem}

\newcommand{\fig}[1]{Fig.~\ref{#1}}
\newcommand{\Fig}[1]{Figure~\ref{#1}}
\newcommand{\Tab}[1]{Table~\ref{#1}}
\newcommand{\eq}[1]{Eq.~(\ref{#1})}

\newcommand{\Eq}[1]{Equation~(\ref{#1})}

\newcommand{\nT}{n(\mathbf{r})}
\newcommand{\nTbar}{n^\mathrm{trap}}
\newcommand{\VT}{V^\mathrm{trap}}
\newcommand{\betaT}{\beta^\mathrm{trap}}
\newcommand{\omegab}{\overline{\omega}}
\newcommand{\partialfix}[3]{ \left.\frac{\partial #1}{\partial #2}\right|_{#3} }

\newcommand{\ntrap}{n^\mathrm{trap}}
\newcommand{\Tr}{\mathrm{Tr}}
\newcommand{\expectation}[1]{\langle #1 \rangle}
\newcommand{\bkt}[1]{\left({#1}\right)}

\begin{document}

\preprint{APS/123-QED}

\title{Temperature-Dependent Contact of Weakly Interacting Single-Component Fermi Gases and Loss Rate of Degenerate Polar Molecules}

\author{Xin-Yuan Gao}
\affiliation{%
Department of Physics, The Chinese University of Hong Kong, Shatin, New Territories, Hong Kong, China
}%
\author{D. Blume}
\affiliation{%
Homer L. Dodge Department of Physics and Astronomy, The University of Oklahoma, 440 W. Brooks Street, Norman, Oklahoma 73019, USA
}%
\affiliation{
Center for Quantum Research and Technology, The University of Oklahoma, 440 W. Brooks Street, Norman, Oklahoma 73019, USA
}
\author{Yangqian Yan}%
 \email{yqyan@cuhk.edu.hk}
\affiliation{%
Department of Physics, The Chinese University of Hong Kong, Shatin, New Territories, Hong Kong, China
}
\affiliation{
The Chinese University of Hong Kong Shenzhen Research Institute, 518057 Shenzhen, China
}%

\date{\today}

\begin{abstract}

Motivated by the experimental realization of single-component degenerate Fermi gases of polar ground state $\mathrm{KRb}$  molecules with intrinsic two-body losses [L. De Marco, G. Valtolina, K. Matsuda, W. G. Tobias, J. P. Covey, and J. Ye, A degenerate Fermi gas of polar molecules, \href{https://www.science.org/doi/10.1126/science.aau7230}{Science \textbf{363}, 853 (2019)}], this work studies the finite-temperature loss rate of single-component Fermi gases with weak interactions.
First, we establish a relationship between the two-body loss rate and the $p$-wave contact.
Second, we evaluate the contact of the homogeneous system in the low-temperature regime using $p$-wave Fermi liquid theory and in the high-temperature regime using the second-order virial expansion.
Third, conjecturing that there are no phase transitions between the two temperature regimes, we smoothly interpolate the results to intermediate temperatures. 
It is found that the contact is constant at temperatures close to zero and increases first quadratically with increasing temperature and finally---in agreement with the Bethe-Wigner threshold law---linearly at high temperatures. 
Fourth, applying the local-density approximation, we obtain the loss-rate coefficient for the harmonically trapped system, reproducing the experimental KRb loss measurements within a unified theoretical framework over a wide temperature regime without fitting parameters. 
Our results for the contact are not only applicable to molecular $p$-wave gases but also to atomic single-component Fermi gases, such as ${}^{40}\text{K}$ and ${}^6\text{Li}$.
\end{abstract}

\maketitle

\textit{Introduction.---} Ultracold polar molecules possess, thanks to their highly controllable long-range anisotropic interactions, a dipole moment. 
This, coupled with their ro-vibrational degrees of freedom, make them promising candidates for quantum computation~\cite{zadoyan2001manipulation,demille2002quantum,andre2006coherent,rabl2006hybrid} and quantum simulations~\cite{carr2009cold,osterloh2007strongly,buchler2007strongly,gorshkov2011tunable,baranov2012condensed}. 
For example, stable degenerate molecular gases can be used to simulate strongly interacting lattice spin models, providing a promising platform for the study of anyonic excitations~\cite{micheli2006toolbox}. 
Even though degenerate atomic Bose and Fermi gases are now being produced routinely, the experimental realization of degenerate molecular gases had been, despite impressive progress by many groups~\cite{bohn2017cold,hummon20132d,barry2014magnetooptical,anderegg2017radio,truppe2017molecules,anderegg2018laser,ni2008high, park2015ultracold, seesselberg2018modeling, takekoshi2014ultracold, molony2014creation, guo2016creation, rvachov2017longlived}, up until 2019~\cite{demarco2019degenerate} hampered detrimentally by losses. 
The experimentally measured loss-rate coefficients were found to increase linearly with the temperature, in nice agreement with predictions derived from two-body physics, i.e., the Bethe-Wigner threshold law~\cite{wigner1948behavior,bethe1935theory,sadeghpour2000collisions,quemener2010strong} and multi-channel quantum defect theory~\cite{greene1982general,idziaszek2010universal}.

In 2019, experimentalists realized molecular $\mathrm{{}^{40}K{}^{87}Rb}$ gases with temperatures $T$ as low as $T/T_F \approx 0.3$, where $T_F$ denotes the Fermi temperature. In a single-component system, the dominant scattering channel is $p$-wave since $s$-wave collisions are prohibited by the Pauli exclusion principle. Importantly, chemical reactions were found to be suppressed in the quantum degenerate regime, i.e., the loss-rate coefficient was found to deviate from the linear temperature dependence observed at higher temperatures~\cite{demarco2019degenerate}. 
Several groups have attempted to explain this intriguing behavior~\cite{he2020exploring,he2020universal}.: Ref.~\cite{he2020exploring} used the master equation, considered fermionic statistics, intermediate four-body complexes, and other factors while Ref.~\cite{he2020universal} connected two-body losses to three distinct $p$-wave contacts.
Yet, a robust theoretical formulation that yields convincing agreement with the experimental data over the entire temperature regime is still lacking.
What role do quantum statistics and many-body effects play as the temperature drops? 
Can the losses be linked to microscopic few-body parameters?
And, if so, how can this be accomplished? 

This letter theoretically investigates, starting with a non-Hermitian Hamiltonian with $p$-wave zero-range interactions, the loss-rate coefficient and arrives at predictions that agree well with published experimental observations. Our key findings are: 
(I) Explicit expressions for the temperature-dependent $p$-wave contact of the weakly-interacting $p$-wave gas are obtained in terms of the low-energy two-body scattering parameters. 
(II) We show that the dominant contribution to the loss rate is---for the experimentally accessible temperature regime---proportional to the $p$-wave contact~\cite{yu2015universal,luciuk2016evidence,yoshida2015universal,zhang2017contact,he2016concept}, defined from the $p$-wave scattering volume; specifically, effective-range contributions are negligible in the weakly interacting regime considered. 
(III) The low- and high-temperature regimes are described by Fermi liquid theory and virial expansion, respectively. 
Motivated by the conjectured absence of a phase transition at intermediate temperatures, the low- and high-temperature predictions are interpolated smoothly.
(IV) Using the local-density approximation, we calculate the loss-rate coefficient of harmonically trapped systems following the experimental procedure~\cite{demarco2019degenerate}. 
The resulting loss curve depends on a single atomic physics parameter, namely the imaginary part of the $p$-wave scattering volume, which is known from experimental high-temperature data~\cite{ospelkaus2010quantumstate} and multi-channel quantum defect theory calculations~\cite{idziaszek2010universal,he2020exploring}.
Our parameter-free theory predictions are in excellent agreement with the experimental data from 2019.

\textit{Effective Hamiltonian and loss relation.---}
For reactive molecular systems such as $\mathrm{KRb}$, two-body losses are triggered by chemical reactions: Two incoming molecules collide and form a four-body complex $\mathrm{K_2Rb_2}$. 
Subsequently, the four-body complex breaks up into $\mathrm{K_2}$ and $\mathrm{Rb_2}$ molecules. Since this process converts internal energy to kinetic energy, the $\mathrm{K_2}$ and $\mathrm{Rb_2}$ molecules become untrapped and fly away. 
It is important to note that the intermediate four-body complexes are transient, with lifetimes of the order of $\mathrm{ps}$ at high temperatures~\cite{bauer1979four,miller1972molecular}. Even in recent lifetime-enhancing experimental set-ups~\cite{hu2019direct,liu2020photoexcitation}, the formation and decay of the four-body complexes is still of the order of $\mathrm{\mu s}$, which is significantly faster than the $\mathrm{ms}$ time scale of interest to us. 
Thus, we treat the chemical reaction as a non-Hermitian loss process and model the system by an effective non-Hermitian Hamiltonian $\hat{H}_{\text{eff}}$, which is equivalent to a Lindblad master equation formulation where the jump operator $L$ contains a term proportional to $\Psi(\mathbf{r'})\Psi(\mathbf{r})$, annihilating two molecules when their positions $\mathbf{r'}$ and $\mathbf{r}$ are close to each other~\cite{braaten2017lindblad}
($\Psi(\mathbf{r})$ is the fermionic field operator).
Assuming a homogeneous system, the effective Hamiltonian reads~\cite{SM}
    \begin{equation}
    \begin{split}
        &\hat{H}_{\mathrm{eff}}=\int \mathrm{d}^3\mathbf{r}\Psi^\dagger(\mathbf{r})\left(-\frac{\hbar^2\nabla_\mathbf{r}^2}{2 m}\right)\Psi(\mathbf{r})\\
        &+\frac{1}{2}\int\mathrm{d}^3\mathbf{r}\mathrm{d}^3\mathbf{r'}\Psi^\dagger(\mathbf{r})\Psi^\dagger(\mathbf{r'})U(|\mathbf{r}-\mathbf{r'}|)\Psi(\mathbf{r'})\Psi(\mathbf{r}),
    \end{split}
    \label{Hamiltonian}
\end{equation}
where $U$ denotes the effective isotropic two-body interaction. 
Each fermionic KRb molecule is treated as a ``fundamental unit" and $U$ describes collisions between two KRb molecules. 
The non-Hermiticity of $\hat{H}_{\text{eff}}$ comes from the imaginary part of $U$, which models losses due to K$_2$ and Rb$_2$ leaving the trap. 
Defining the particle number operator $\hat{N}=\int\mathrm{d}^3\mathbf{r}\Psi^\dagger(\mathbf{r})\Psi(\mathbf{r})$, the particle loss $\mathrm{d}N/\mathrm{d}t$ can be related to the imaginary part of the effective Hamiltonian~\cite{SM},  
$
    \frac{\mathrm{d}N}{\mathrm{d}t}=\frac{4}{\hbar}\Im\langle \hat{H}_\mathrm{eff}\rangle,
    \label{losseq}
$
where the average particle number $N$ is given by $\langle\hat{N}\rangle$, with $\langle \cdot \rangle$ denoting the trace.

Without an external electric field, the interaction $U$ between two molecules is of van der Waals type and short ranged. In the low-energy scattering regime, $U(|\mathbf{r}-\mathbf{r'}|)$ can be parameterized by the complex scattering volume $v_p$, which appears in the leading order term of the low-energy expansion of the $p$-wave phase shift $\delta_p(k)$, $\tan(\delta_p(k))=-v_pk^3$. 
In what follows, we concentrate on the weak-interaction and weak-reaction regimes, in which both the real and imaginary parts of $v_p$ are significantly smaller than $1/n=V/N$, where $n$ and $V$ denote the density and volume, respectively.  
In this limit, we find~\cite{SM} that the loss rate can be expressed in terms of the scattering volume 
$v_p$ and the $p$-wave contact $C_v$ conjugate to $v_p$,
\begin{equation}
    \frac{\mathrm{d}N}{\mathrm{d}t}=\frac{6\hbar}{m}C_v\frac{\Im(v_p)}{(\Re(v_p))^2};
    \label{contacteq}
\end{equation}
here, $C_v=-\frac{2m}{3\hbar^2}\frac{\partial F(\Re(v_p))}{\partial v_p^{-1}}$~\cite{yu2015universal,luciuk2016evidence,yoshida2015universal,zhang2017contact,he2016concept} and $F$ denotes the Helmholtz free energy and $m$ the mass of the Fermi gas constituents.
Since the contact $C_v$, which is an extensive variable, is evaluated using the real part of $v_p$, and not the complex $v_p$, $C_v$ is purely real. The physical picture behind the derivation is that the system first thermalizes and that losses are subsequently turned on perturbatively.
Equation~(\ref{contacteq}) is analogous to the universal loss relation for $s$-wave systems, where the loss is proportional to the $s$-wave contact~\cite{braaten2008exact,braaten2013universal,braaten2017lindblad}. 

\textit{Effective range.---}
The low-energy properties of strongly-interacting $p$-wave systems do not only depend on $v_p$ but also on the effective range $R$~\cite{yu2015universal}, which appears at sub-leading order in the low-energy expansion, $\tan(\delta_p(k))=-v_pk^3+R^{-1}v_p^2 k^5$. 
Assuming the naturalness of the expansion in the weakly interacting regime (this means that $v_p$ and $R$ are, respectively, proportional to $l_0^3$ and $l_0$, where $l_0$ is the characteristic length of the interaction), the effective-range contribution to the thermodynamics can be neglected.
This follows because  $|R^{-1}v_p^2| k^5$ is a factor of $l_0^2k^2$ smaller than $|v_p|k^3$. Setting $l_0$ equal to the van der Waals length of KRb molecules ($l_0\simeq118a_0$)~\cite{idziaszek2010universal}, the quantity $l_0^2k^2$, in turn, can be shown to be much smaller than 1 for the experimental temperature of several hundreds $\mathrm{n K}$~\cite{demarco2019degenerate}.
The above argument can be extended to justify dropping contributions from other higher-order terms in the expansion of the phase shift, which may, e.g., arise from the van der Waals tail of the interaction~\cite{levy1963low,zhang2010scattering}.

Our conclusion is consistent with the literature. 
First, at zero temperature, the leading-order contribution to the energy---calculated within the Fermi liquid theory---does not depend on the effective range~\cite{ding2019fermiliquid}. 
Second, at high temperatures, the loss-rate coefficient scales linearly with $\Im{(v_p)}T$. 
Third, the fact that a $T^2$ scaling, which arises from effective-range contributions based on dimensional arguments, is not observed experimentally~\cite{ospelkaus2010quantumstate} additionally supports that effective-range contributions play a minor role.
\begin{figure}[t]
    \centering
    \includegraphics[width=0.49\textwidth]{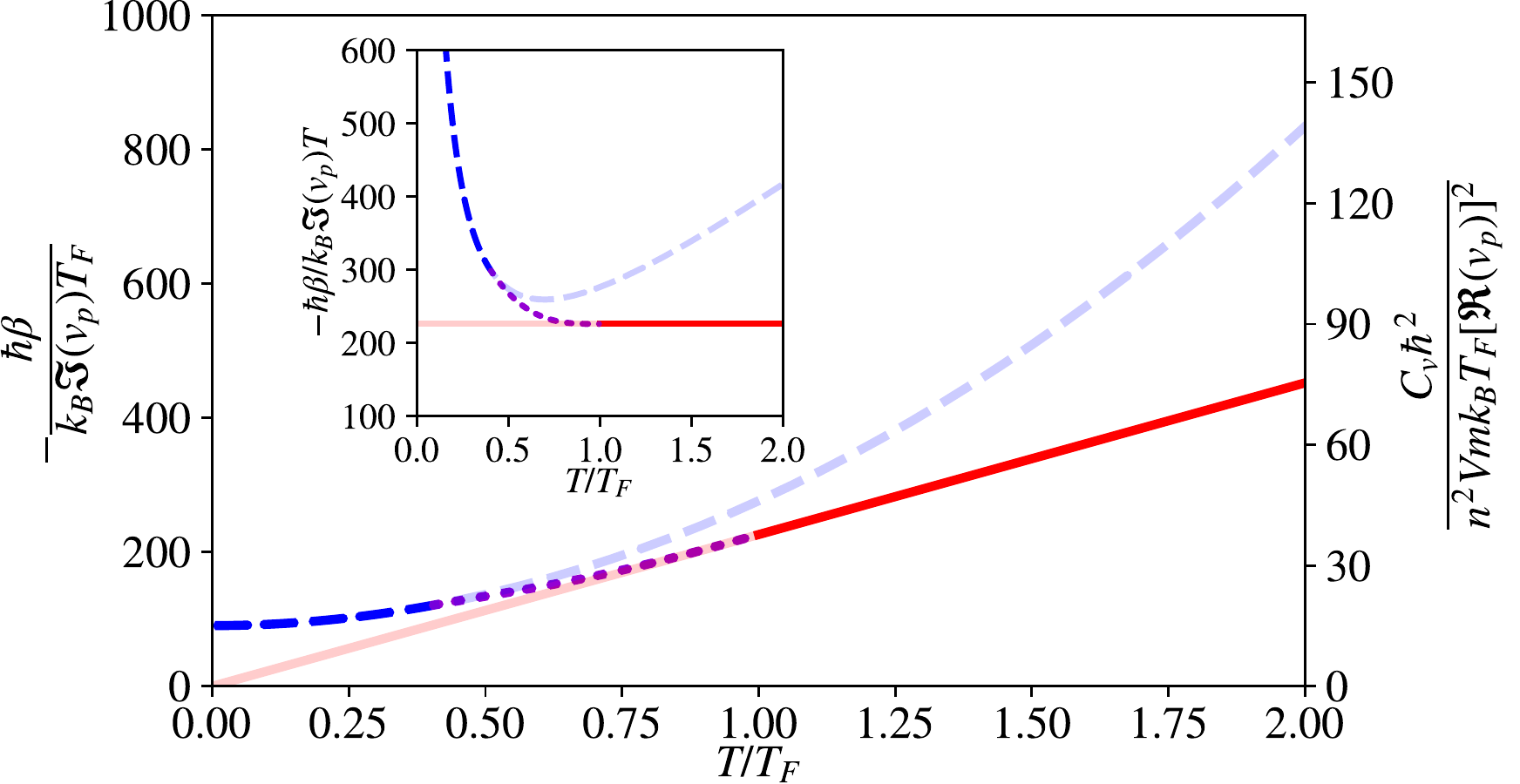}
    \caption{%
    The loss-rate coefficient $\beta$ (left axis) and contact $C_v$ (right axis) as a function of the temperature $T$. Blue dashed and red solid lines are obtained from the Fermi liquid theory and the second-order virial expansion, respectively. Transparent lines extend the predictions beyond their range of validity. The purple dotted line interpolates the low-$T$ and high-$T$ results.  Inset: Same data but showing $\beta/T$ instead of $\beta$.}
    \label{homo_loss}
\end{figure}

\textit{Homogeneous system: high temperature.---} 
In the $T/T_F \gtrsim 1$ regime, we adopt the virial expansion that has been used extensively for $s$-wave systems~\cite{ho2004high,liu2013virial}.
The grand potential $\Omega$ is expanded up to the second order in the fugacity $z$.
The second virial coefficient is obtained by the Beth-Uhlenbeck formalism~\cite{beth1937quantum} with the bound state contribution discarded.
This is consistent with our assumption that the formation of bi-molecular complexes leads to losses that are accounted for by the complex phase shift.
In the weak-interaction limit, $n\Re(v_p)\ll 1$, we obtain 
\begin{equation}
    C_v=\dfrac{12\pi m k_B T n^2 V (\Re(v_p))^2}{\hbar^2}\label{virialCv}.
\end{equation}
Defining the loss-rate coefficient $\beta$ through $({\mathrm{d}n}/{\mathrm{d}t})=-\beta n^2$,
we find the following from Eqs.~(\ref{contacteq}) and (\ref{virialCv}):
\begin{equation}
    \beta=-\frac{72 \pi k_B}{\hbar}T\Im(v_p).
\label{highTbeta}
\end{equation}
It can be seen that $\beta$ is proportional to $T$ as predicted by the Bethe-Wigner threshold law and previous high-temperature results~\cite{he2020universal}.
The solid lines in the main part and inset of \fig{homo_loss} show $C_v$, $\beta$, and $\beta/T$, as predicted by Eqs.~(\ref{virialCv}) and (\ref{highTbeta}), as a function of the temperature.

\textit{Homogeneous system: low temperature.---} In the $T/ T_F \ll 1$ regime, the thermodynamics of $p$-wave gases is expected to be governed by many-body effects. 
To account for interactions and Fermi statistics, we apply $p$-wave Fermi liquid theory~\cite{baym2008landau,ding2019fermiliquid}. 
Working, as above, at leading-order in  $n\Re(v_p)$, we obtain 
\begin{equation}
    \begin{split}
            C_v=&\dfrac{12\times6^{2/3}\pi^{7/3} n^{8/3} V (\Re(v_p))^2}{5}\\
            &+\dfrac{2^{1/3}\pi^{5/3}m^2k_B^2T^2n^{4/3} V (\Re(v_p))^2}{3^{2/3}\hbar^4}.
    \end{split}
    \label{CvFermi}
\end{equation}
The two-body contact $C_v$ contains terms that are proportional to $n^{8/3}$ and $n^{4/3}$. 
This is distinct from the high-temperature regime, where $C_v$ is proportional to $n^2$. 
Since the Fermi temperature $T_F=\frac{6^{2/3}\pi^{4/3}\hbar^2n^{2/3}}{2m k_B}$ is proportional to $(n^{4/3})^{1/2}$, the loss-rate coefficient becomes
\begin{equation}
    \beta=-\frac{144 \pi k_B}{5\hbar}T_F\Im(v_p)-\frac{6\pi^3 k_B}{\hbar}\frac{T^2}{T_F}\Im(v_p), 
    \label{lowTbeta}
\end{equation}
i.e., $\beta$ contains a term that is proportional to $T^0$ and a term that is proportional to $T^2$.
This behavior should be contrasted with the linear $\beta \propto T$ scaling for $T/T_F \gg 1$. 
In the low-temperature regime, the many-body energy scale $k_BT_F$ governs the scaling of $C_v$ with $n$ and, correspondingly, the scaling of $\beta$ with $T$. Specifically, the first constant term on the right hand sides of Eqs.~(\ref{CvFermi}-\ref{lowTbeta}) arises from the interacting ground state.  
The second term on the right-hand sides of Eqs.~(\ref{CvFermi}-\ref{lowTbeta}) arises from excitations out of the ground state.
The dashed line in \fig{homo_loss} shows $\beta$, Eq.~(\ref{lowTbeta}), as a function of the temperature. 
It can be seen that the $T^2$ term has a vanishingly small contribution for $T/T_F \lesssim 0.3$. 
Correspondingly, $\beta/T$ increases for $T/T_F \ll 1$ with decreasing $T$ (see the inset of \fig{homo_loss}),
i.e., \eq{lowTbeta} predicts an enhancement rather than a suppression.

\textit{Homogeneous system: intermediate temperature.---} 
We now prove that a weakly-interacting $p$-wave Fermi gas can only have one phase transition at extremely low temperatures, thereby justifying interpolation of the
 low- and high-temperature expressions in the intermediate  $0.4<T/T_F<1$ regime.
Since all the Lee-Yang zeros (LYZ)~\cite{yang1952statistical,lee1952statistical} of the non-interacting single-component Fermi gas except for the LYZ at complex infinity lie on the negative real axis away from the origin~\cite{huang2008statistical}, only the LYZ at complex infinity may move to the positive real axis as interactions are turned on perturbatively.
If $\Re(v_p)$ and $\Re(R)$ have opposite signs, there exists no weakly-bound two-body state below the scattering threshold, indicating that the hypothetical transition would be to a BCS phase.
Since the BCS transition temperature $T_c$ scales as $T_F\exp(\frac{\pi}{2 k_F \Re(R)}-\frac{\pi}{2 k_F^3 |\Re(v_p)|})$~\cite{ho2005fermion,iskin2006evolution,yao2018normalstate}, $T_c$ would be several orders of magnitude smaller than $T_F$ and thus below the temperature of interest in this work.
If, on the other hand, $\Re(v_p)$ and $\Re(R)$ have the same sign, the two-body bound state energy $E_b=\hbar^2\Re(R)/m\Re(v_p)$ is positive, indicating that the hypothetical transition would be to a BEC phase.
Noting that $E_b$ scales as ${\hbar^2}/{m l_0^2}$, which is much larger than the typical system energy ${\hbar^2k^2}/{m}$, the gas prepared experimentally should not support such bound states, implying that the system is far from the BEC transition. 
Note that our arguments do not rule out the existence of other more exotic phases.
Motivated by the absence of BCS- and BEC-phase transitions in the intermediate temperature regime, we conjecture that the low- and high-temperature curves provide upper and lower bounds for $\beta(T)$ for $0.4<T/T_F<1$. 
Since the low- and high-temperature expressions differ by only 15~\% at $T/T_F=0.7$, the interpolation (the dotted lines in Fig.~\ref{homo_loss}) is expected to be quite accurate.

\begin{figure}[t]
    \centering
    \includegraphics[width=0.49\textwidth]{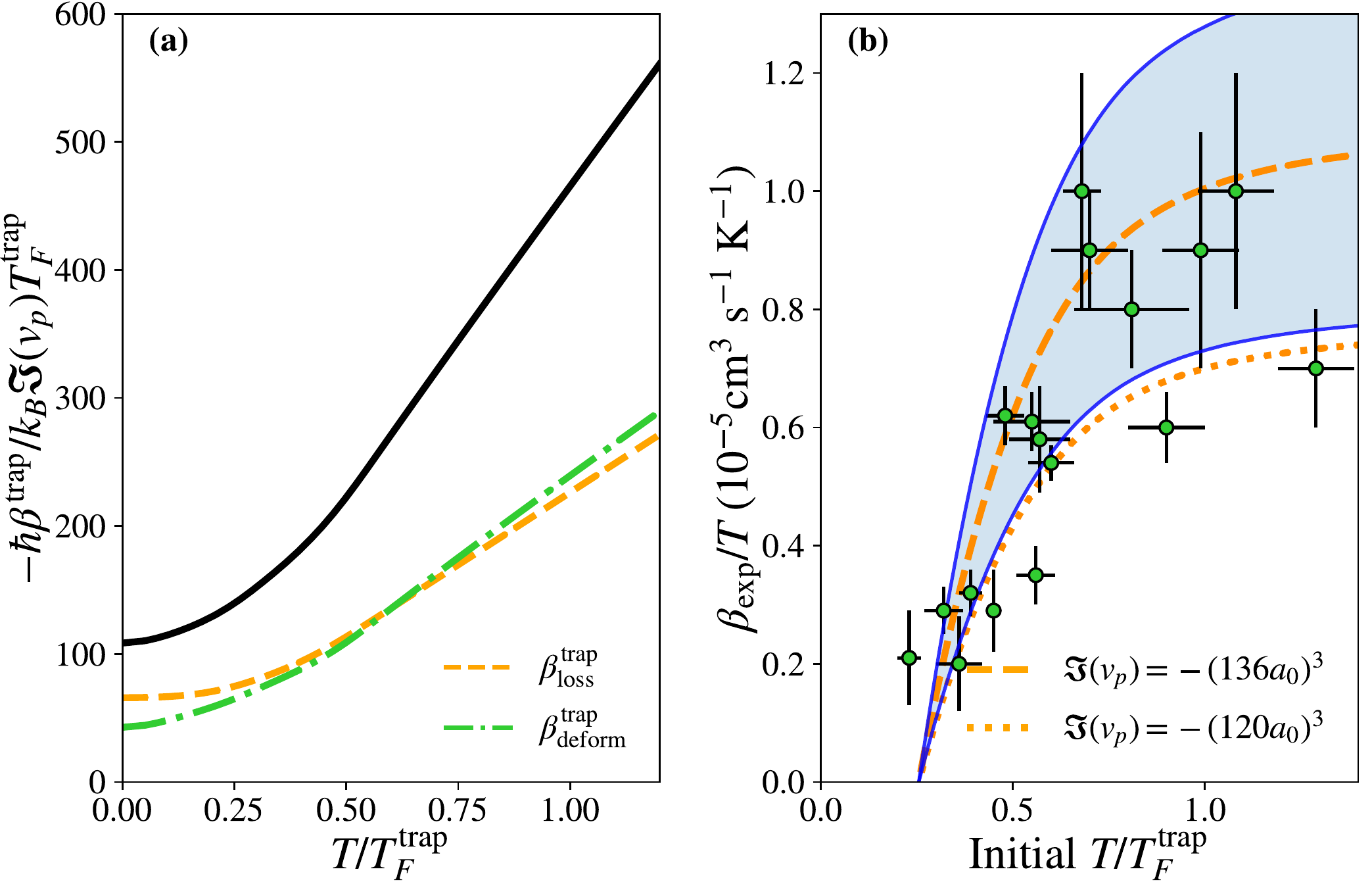}
    \caption{ (a) The black solid, orange dashed, and green dash-dotted lines show $\betaT$, $\beta_{\text{loss}}^{\text{trap}}$, and $\beta_{\text{deform}}^{\text{trap}}$ against $T$. (b) Theory predictions for $\beta_{\text{exp}}$ (lines) are compared with measurements~\cite{demarco2019degenerate,demarco2019replicationa} (symbols) as a function of the initial temperature $T$. The dashed curve with the blue shaded region and the dotted curve are for $\Im(v_p)=-(136^{+11}_{-14}a_0)^3$ (extracted from~\cite{ospelkaus2010quantumstate}) and $\Im(v_p)=-(120a_0)^3$~\cite{idziaszek2010universal}.
    }
    \label{trap_and_exp}
\end{figure}

\textit{Loss in harmonically trapped system.---}
The most straightforward definition of the loss-rate coefficient of an inhomogeneous system is through a local quantity $\beta(\mathbf{r})$, defined by replacing the density $n$ by the local density $n(\mathbf{r})$. 
Application of this generalization to the experiment requires spatially resolved loss measurements, which may not be possible due to, e.g., limited resolution caused by the finite waist of imaging lasers.
To derive broadly applicable results, we instead calculate the global loss rate by applying the local-density approximation to our results for the homogeneous system.
The global loss-rate coefficient $\beta^{\text{trap}}$ for the trapped system is defined through $\frac{\mathrm{d}\nTbar}{\mathrm{d}t}=-\betaT (\nTbar)^2$,
where the average in-situ density $\nTbar=N/V^{\text{trap}}$ depends on the trap volume $V^{\text{trap}}=N^2 \left[\int\mathrm{d}^3\mathbf{r}(\nT)^2\right]^{-1}$.
Evaluating $\mathrm{d}N/\mathrm{d}V^{\text{trap}}$,
$\betaT$ naturally separates into a sum of
two quantities, namely $\beta^{\text{trap}}_{\text{loss}}$ and $\beta^{\text{trap}}_{\text{deform}}$, which originate from the time derivative of $N$ and of $(V^{\text{trap}})^{-1}$, respectively.
We find~\cite{SM}
\begin{eqnarray}
 \beta_\text{loss}^\text{trap} = -\frac{\VT}{N^2}\frac{\mathrm{d}N}{\mathrm{d}t}=\frac{\int\mathrm{d}^3 r \beta(\mathbf{r})(\nT)^2}{\int\mathrm{d}^3 r (\nT)^2} \label{betaTr}
\end{eqnarray}
and
\begin{equation}
    \begin{split}
        &\beta^{\text{trap}}_{\text{deform}}=
\frac{1}{N}\frac{\mathrm{d}\VT}{\mathrm{d}t}\\
&=\dfrac{2\left(\int\mathrm{d}^3 r n(\mathbf{r})\right)\left(\int\mathrm{d}^3 r \beta(\mathbf{r})[n(\mathbf{r})]^3\right)}{\left(\int\mathrm{d}^3 r[n(\mathbf{r})]^2\right)^2}-\dfrac{2\int\mathrm{d}^3 r \beta(\mathbf{r})[n(\mathbf{r})]^2}{\int\mathrm{d}^3r[n(\mathbf{r})]^2}.\\
    \end{split}
    \label{betaTd}
\end{equation}
In Eqs.~(\ref{betaTr}) and (\ref{betaTd}), $\beta(\mathbf{r})$ is the local loss-rate coefficient, evaluated at each $\mathbf{r}$ using the properties of the homogeneous system with 
$k_B T_F=k_B T_F(\mathbf{r})=\frac{\hbar^2}{2m} (6\pi^2\nT)^{2/3}$. 
If $T \gg T_F(\mathbf{0})$, $\beta(\mathbf{r})$ follows \eq{highTbeta} closely for all $\mathbf{r}$. 
Since \eq{highTbeta} has no dependence on $\mathbf{r}$, $\beta(\mathbf{r})$ can be pulled out of the integrand and $\beta_{\text{loss}}^{\text{trap}}= \beta$.  
At lower temperatures, in contrast, 
$\beta_\text{loss}^\text{trap}$ receives---as discussed in more detail below--- contributions from the homogeneous $\beta$ for a range of different densities.

The solid line in \fig{trap_and_exp}(a) shows $\beta/T_F^\mathrm{trap}$ as a function of $T/T_F^\mathrm{trap}$, where $T_F^\mathrm{trap}=(6N)^{1/3}\hbar\bar{\omega}/k_B$ is the global Fermi temperature of the trapped system and $\bar{\omega}$ is the mean angular trap frequency. 
The dashed and dash-dotted lines show $\beta_\text{loss}^\text{trap}$ and $\beta_{\text{deform}}^{\text{trap}}$, respectively. 
It can be seen that these two terms contribute approximately equally for all temperatures considered.
Even though $\beta_{\text{deform}}^{\text{trap}}$ contributes to $\beta^{\text{trap}}$, we emphasize that it characterizes the change of the trap volume (or cloud size) with time and not directly the change of the number of trapped particles with time. 
If $\frac{\mathrm{d}\nT}{\mathrm{d}t}\propto (\nT)^\alpha$ with $\alpha>1$, the density at the center of the cloud decreases faster than the density at the edge, causing the trap volume to increase with time.  

Since both $\beta_\text{loss}^\text{trap}$ and $\beta_{\text{deform}}^{\text{trap}}$ contribute to the global loss-rate coefficient $\beta^{\text{trap}}$, experimentalists need to measure---if the gas is, as in the recent groundbreaking KRb experiment~\cite{demarco2019degenerate}, prepared with an inhomogeneous density---both $\mathrm{d}N/\mathrm{d}t$ and $\mathrm{d}V^{\text{trap}}/\mathrm{d}t$. To facilitate the analysis of the experimental data, Ref.~\cite{demarco2019degenerate} approximated $\nTbar$ by the semi-classical expression $\nTbar_\mathrm{exp}(T)=N\bar\omega^3(m/\pi k_B T)^{3/2}/8$; in addition, $T$ in $\nTbar_\mathrm{exp}$ was approximated by $T_\mathrm{exp}$. Both the temperature $T_\mathrm{exp}$ and particle number $N$  were determined through ballistic expansion. 
With $T_\mathrm{exp}$ and $N$ measured, $\mathrm{d}V^{\text{trap}}/\mathrm{d}t$ was evaluated using $(\mathrm{d}V^{\text{trap}}/\mathrm{d}T_{\text{exp}})(\mathrm{d}T_{\text{exp}}/\mathrm{d}t)$.
Finally, a loss-rate coefficient was extracted through fits. 
While Ref.~\cite{demarco2019degenerate}'s goal was to extract the loss-rate coefficient $\beta_{\text{loss}}^{\text{trap}}$, we refer to the experimentally extracted loss-rate coefficient as $\beta_{\text{exp}}$ to remind the reader that the result relies on the use of $n_{\text{exp}}^{\text{trap}}(T_{\text{exp}})$.

The scheme described above is valid at $T\gtrsim T_F^\mathrm{trap}$, where the identities $\nTbar_\mathrm{exp}(T_\mathrm{exp})=\nTbar(T)$ and $\beta=\beta_\text{loss}^\text{trap}=\beta_{\text{exp}}$ hold. 
For $T \lesssim T_F^\mathrm{trap}$, however, the two identities no longer hold. 
For example, we find that $T_\mathrm{exp}$ is twice as large as $T$ for $T/T_F^{\text{trap}}=0.3$; similarly, $n^{\text{trap}}_{\text{exp}}$ deviates from $n^{\text{trap}}$~[see Fig.~(S3) in Supplemental Material~\cite{SM}].
To connect our results to $\beta_{\text{exp}}$, we follow a two-step approach. 
First, using \eq{highTbeta}, we extract $\Im(v_p)=-(136^{+11}_{-14}a_0)^3$  from experimental high-temperature data~\cite{ospelkaus2010quantumstate},  where $a_0$ is the Bohr radius.
This is comparable to the value $-(120a_0)^3$ obtained by multi-channel quantum defect theory~\cite{idziaszek2010universal}.
Second, establishing a formal connection between $\beta^{\text{trap}}$ and 
$\beta_\mathrm{exp}$~[see Eqs.~(S112) and (S113) in Supplemental Material~\cite{SM}], we calculate $\beta_\mathrm{exp}$ over the entire temperature regime by emulating the ballistic expansion to get $T_{\text{exp}}$. 
The dashed [dotted] lines in \fig{trap_and_exp}(b) show our results as a function of the \textit{initial} temperature $T/T_F^{\text{trap}}$ for $\Im(v_p)=-(136^{+11}_{-14}a_0)^3$ [$\Im(v_p)=-(120a_0)^3$]. 
Our theory predictions agree within error bars with the experimental data. 

In summary, we theoretically studied the temperature dependence of the contact and loss rate of a homogeneous Fermi gas
with weak $p$-wave interactions parameterized by the complex $p$-wave scattering volume in the normal phase. 
In the quantum degenerate regime, the two-body loss rate is not directly proportional to the temperature, as one would expect naively by extrapolating the high-temperature results to below the Fermi temperature, but governed by two terms: one term that is directly proportional to $T^2$ and another term that is independent of $T$. 
Applying the local-density approximation, our parameter-free loss-rate coefficient predictions for the harmonically trapped system are found to agree with experimental measurements for a molecular KRb gas over the entire temperature regime considered. 
In addition, our theory framework identifies the physical origin of the leading-order behaviors within a transparent unified picture, resolving the important---yet puzzling---experimental observation that the $\beta \propto T$ scaling law does not hold below the Fermi temperature.

Our work is expected to stimulate further theoretical and experimental works at finite temperatures near the Fermi temperature.
For example, our theory study lays the foundation for developing robust experimental temperature calibration schemes. 
We emphasize that our theoretical framework and predictions for the contact are universal, i.e., they are also applicable to single-component atomic gases, including $^6\mathrm{Li}$ and $^{40}\mathrm{K}$. Thus, our theory can be further verified using atomic gas experiments.

    We thank Jun Ye and Junru Li for very useful discussions and information. 
    This work is supported by the National Natural Science Foundation of China under Grant No. 12204395 and CUHK Direct Grant No. 4053583 and No. 4053535. 
    D.~B. acknowledges support by the National Science Foundation (NSF) through grant No. PHY-2110158.


%

\clearpage
\widetext
\begin{center}
\textbf{\large Supplemental Material: Temperature-dependent contact of weakly interacting single-component Fermi gases and loss rate of degenerate polar molecules}
\end{center}
\setcounter{equation}{0}
\setcounter{figure}{0}
\setcounter{table}{0}
\setcounter{page}{8}
\makeatletter
\renewcommand{\theequation}{S\arabic{equation}}
\renewcommand{\thefigure}{S\arabic{figure}}
\renewcommand{\bibnumfmt}[1]{[S#1]}

\tableofcontents

~\\

The derivations and discussions in this supplemental material are based on the $p$-wave effective range theory~\cite{bertulani2002effective,braaten2012renormalization}, which expands the effective $p$-wave phase shift $\delta_p(k)$ in terms of scattering volume $v_p$ and effective range $R$,
\begin{equation}
    \frac{\tan(\delta_p(k))}{k^3}=-v_p+\frac{v_p^2}{R}k^2.
    \label{SM::phaseshift}
\end{equation}
Here, $k$ is the relative wave vector.
We remind readers that we write $v_p=\Re(v_p)+\imath \Im(v_p)$ and similarly for $R$, i.e., the imaginary parts of $v_p$ and $R$ are $+\Im(v_p)$ and $+\Im(R)$, respectively. This convention differs some previous works~\cite{hutson2007feshbach,idziaszek2010universal}.
There are two reasons why the effective range term is explicitly accounted for in the following calculations, despite the fact that we argue in the main text that the effective range does not contribute at the temperatures where the experiments are conducted. 
First, it is critical for some intermediate steps. 
If $R$ was dropped at the outset, some integrals would diverge. 
Second, Sec.~\ref{SMsect::virial} derives contributions from $R$ and verifies our assertion presented in the main text that $R$ leads to a $T^2$ term in the loss-rate coefficient. 
The results also allow us to connect to a conjecture made in Ref.~\cite{he2020universal}.

In the main text, in order to discuss the negligible contribution of $R$ to the thermodynamics, we assume the naturalness of $v_p$ and $R$, i.e., except for the dimensionful parts, which inherit the length scale $l_0$ of the interaction, their dimensionless prefactor should of order 1. 
Because we assume that $n |v_p|$ is much smaller than 1, the characteristic length $l_0$ of the interaction $U$ between particles is much smaller than the interparticle spacing $n^{1/3}$. 
Thus, when $n|v_p|$ goes to zero,  the quantity $n^{1/3}|R|$ also goes to zero. This naturalness argument is used extensively in the discussions below.

In summary, in both the main text and this supplemental material, our calculations apply provided
\begin{equation}
    n^{5/3}\left|\frac{v_p^2}{R}\right| \ll n \left|v_p\right| \ll 1.
    \label{SM::inequality}
\end{equation}
The two inequalities follow from two assumptions: (1) the naturalness of $v_p$ and $R$; (2) $n |v_p|\ll 1$. 
For ground state $\mathrm{{}^{40}K{}^{87}Rb}$ molecules, which are our main interest, the naturalness of $v_p$ and (2) can be directly shown. 
Both the real and imaginary parts of $v_p$ have been predicted to be about $-1.064 l_0^3$~\cite{idziaszek2010universal}, where $l_0=118a_0$ ($a_0$ is the Bohr radius) is the characteristic length of the van der Waals potential. 
In the experiment, the typical (average in-situ) density of the system is $\sim10^{12}\mathrm{cm^{-3}}$; thus $n|v_p|\sim3.7\times10^{-7}\ll1$. The effective range $R$ of $\mathrm{{}^{40}K{}^{87}Rb}$
has, to the best of our knowledge, neither been calculated nor measured. 
Nevertheless, measurements of $R$ for other atomic ultracold gases provide an indirect argument for verifying the hypothesis. 
For ground state $\mathrm{{}^{40}K}$ atoms, $v_p\simeq(96.74a_0)^3$ and $R\simeq46.22a_0$~\cite{ticknor2004multiplet,luciuk2016evidence}. 
If one treats $|v_p|^{1/3}$ as the length scale of the potential, the coefficient of $R$ is $\sim0.48$, which is of order 1.

The main text as well as Secs.~\ref{SMsect::LDA} and \ref{SMsect::experiment} of this supplemental material use a variety of
 temperatures, densities, and loss-rate coefficients.
 The notation is summarized in \Tab{SM::notationlist}.
\begin{table}
\centering
\begin{tabular}{ |p{1.5cm}||p{13.5cm}| }
 \hline
 \multicolumn{2}{|c|}{Notation List} \\
 \hline
 Symbol         & Description \\
 \hline
 $N$ & total number of particles     \\
 $V$ & volume (homogeneous system)    \\
 $n$ & density (homogeneous system) $n=N/V$    \\
 $T_F$ & Fermi temperature (homogeneous system) $T_F=\frac{6^{2/3}\pi^{4/3}\hbar^2 n^{2/3}}{2 m k_B}$ \\
 $\beta$ & two-body loss-rate coefficient (homogeneous system) $\frac{\mathrm{d}n}{\mathrm{d}t}=-\beta n^2$    \\
 $n(\mathbf{r})$ & local density at $\mathbf{r}$ (harmonically trapped system)\\
 $\ntrap$ & average in-situ density (harmonically trapped system) $\ntrap=\frac{1}{N}\int\mathrm{d}^3\mathbf{r} (n(\mathbf{r}))^2$\\
 $\VT$ & in-situ volume (harmonically trapped system) $\VT=N/\ntrap$\\
 $T_F(\mathbf{r})$ & local Fermi temperature (harmonically trapped system) $T_F(\mathbf{r})=\frac{6^{2/3}\pi^{4/3}\hbar^2 (n(\mathbf{r}))^{2/3}}{2 m k_B}$\\
 $\beta(\mathbf{r})$ & local two-body loss-rate coefficient (harmonically trapped system) $\frac{\mathrm{d}n(\mathbf{r})}{\mathrm{d}t}=-\beta(\mathbf{r}) (n(\mathbf{r}))^2$\\
 $\betaT$ & in-situ two-body loss-rate coefficient (harmonically trapped system) $\frac{\mathrm{d}\ntrap}{\mathrm{d}t}=-\betaT (\ntrap)^2$\\
 $\beta_\text{loss}^\text{trap}$ & component of in-situ two-body loss-rate coefficient that is related to physical loss(harmonically trapped system) Eq.~(\ref{betaTr})\\
 $\beta_\mathrm{deform}^\mathrm{trap}$ & component of in-situ two-body loss-rate coefficient that is related to volume variation (harmonically trapped system) Eq.~(\ref{betaTd})\\
   $T_F^\mathrm{trap}$ & Fermi temperature (harmonically trapped system) $T_F^\mathrm{trap}=(6N)^{1/3}\hbar \omega/k_B$\\
 $T_\mathrm{exp}$ & expansion temperature (experimental measurement) \eq{SM::nrfit} \\
 $\ntrap_\mathrm{exp}$ & in-situ average density (experimental measurement) $\ntrap_\mathrm{exp}(T_\mathrm{exp})=N\omega^3(m/\pi k_B T_\mathrm{exp})^{3/2}/8$\\
 $\VT_\mathrm{exp}$ & in-situ volume (experimental measurement) $\VT_\mathrm{exp}=N/\ntrap_\mathrm{exp}$\\
 $\beta_\mathrm{exp}$ & in-situ two-body loss-rate coefficient (experimental measurement) \eq{SM::conversion} \\
 \hline
\end{tabular}
\caption{Different symbols related to densities, temperatures, and loss-rate coefficients.}
\label{SM::notationlist}
\end{table}

\section{Lindblad equation and loss relation}

For a Hermitian Hamiltonian $\hat{H}$, the density matrix in the Schr\"{o}dinger~picture follows the von Neumann-equation
\begin{equation}
    i\hbar \frac{\partial \rho}{\partial t}=[\hat{H},\rho],
    \label{SM::von1}
\end{equation}
where $\rho=|\psi\rangle\langle\psi|$ is the density matrix and  $|\psi\rangle$ the state vector in the Schr\"{o}dinger~ picture. 
This section shows that the von Neumann-equation takes on a different form for a non-Hermitian Hamiltonian $\hat{H}_{\text{eff}}$.
Using the density matrix framework, we give a detailed deviation of Eq.~(\ref{contacteq}) in the main text.

We start by separating the effective non-Hermitian Hamiltonian $\hat{H}_{\text{eff}}$, Eq.~(\ref{Hamiltonian}), of the homogeneous system into its real and imaginary parts,
\begin{equation}
    \hat{H}_\mathrm{eff}=\Re(\hat{H}_\mathrm{eff})+i\Im(\hat{H}_\mathrm{eff}).
\end{equation}
Using the fermionic field operators $\psi(\mathbf{r})$ and $\psi^{\dagger}(\mathbf{r})$ and accounting for two-body interactions and losses through the complex potential $U$, the real and imaginary parts read
\begin{align}
    \Re(\hat{H}_\mathrm{eff})&=\int\mathrm{d}^3 r\Psi^{\dagger}(\mathbf{r})\left(-\frac{\hbar^2\nabla_{\mathbf{r}}^2}{2 m}\right)\Psi(\mathbf{r})+\frac{1}{2}\int\mathrm{d}^3 r\mathrm{d}^3 r'\Re(U(|\mathbf{r}-\mathbf{r'}|))\Psi^\dagger(\mathbf{r})\Psi^\dagger(\mathbf{r'})\Psi(\mathbf{r'})\Psi(\mathbf{r})\label{SM::ReH}\\
    \Im(\hat{H}_\mathrm{eff})&=\frac{1}{2}\int\mathrm{d}^3r\mathrm{d}^3r'\Im(U(|\mathbf{r}-\mathbf{r'}|))\Psi^\dagger(\mathbf{r})\Psi^\dagger(\mathbf{r'})\Psi(\mathbf{r'})\Psi(\mathbf{r}).
    \label{SM::ImH}
\end{align}
The time-dependent Schr\"{o}dinger~ equation reads
\begin{equation}
    i\hbar\frac{\partial \left|\psi\right \rangle }{\partial t} =(\Re(\hat{H}_\mathrm{eff})+i\Im(\hat{H}_\mathrm{eff}))\left|\psi \right \rangle.
    \label{SM::Schrodinger1}
\end{equation}
Taking the adjoint of \eq{SM::Schrodinger1}, we obtain  
\begin{equation}
    -i\hbar\frac{\partial\left\langle\psi\right| }{\partial t} =\left\langle\psi\right| (\Re(\hat{H}_\mathrm{eff})-i\Im(\hat{H}_\mathrm{eff})).
    \label{SM::Schrodinger2}
\end{equation}
Multiplying \eq{SM::Schrodinger1} from the right with $\left\langle\psi\right|$  and \eq{SM::Schrodinger2} from the left with $\left|\psi \right \rangle$, one obtains
\begin{align}
    &i\hbar\frac{\partial \left|\psi\right \rangle }{\partial t}\left\langle\psi\right| =(\Re(\hat{H}_\mathrm{eff})+i\Im(\hat{H}_\mathrm{eff}))\left|\psi \right \rangle\left\langle\psi\right|,\label{SM::Schrodinger3}\\
    &-i\hbar\left|\psi \right \rangle\frac{\partial\left\langle\psi\right| }{\partial t} =\left|\psi \right \rangle\left\langle\psi\right| (\Re(\hat{H}_\mathrm{eff})-i\Im(\hat{H}_\mathrm{eff})).
    \label{SM::Schrodinger4}
\end{align}
The 
"von Neumann-equation" for the non-Hermitian Hamiltonian $\hat{H}_{\text{eff}}$ is obtained by subtracting \eq{SM::Schrodinger4} from \eq{SM::Schrodinger3}:
\begin{equation}
    i\hbar \frac{\partial \rho}{\partial t}=[\Re(\hat{H}_\mathrm{eff}),\rho]+i\{\Im(\hat{H}_\mathrm{eff}),\rho\}.
    \label{SM::von2}
\end{equation}
\Eq{SM::von2} is a "problematic master equation," because it does not conserve the trace of the density matrix. Specifically, one finds that the change of the trace of $\rho$ with time is governed by the imaginary part of the effective Hamiltonian, 
\begin{equation}
\frac{\partial(\Tr \rho)}{\partial t}=\frac{2\Tr(\Im(\hat{H}_\mathrm{eff})\rho)}{\hbar}.
\end{equation}
To fix the loss of probability, the Lindblad term $-i\sum_n L_n \rho L_n^\dagger$, where $L_n$ is the Lindblad jump operator, can be added to \eq{SM::von2} to change it to a Lindblad equation:
\begin{equation}
    i\hbar \frac{\partial \rho}{\partial t}=[\Re(\hat{H}_\mathrm{eff}),\rho]+i\{\Im(\hat{H}_\mathrm{eff}),\rho\}-i\sum_n L_n \rho L_n^\dagger.
    \label{SM::Lindbladeq}
\end{equation}
The jump operators are determined by requiring that the trace of $\rho$ does not change with time:
\begin{eqnarray}
\frac{\partial(\Tr \rho)}{\partial t}=\frac{2\Tr(\Im(\hat{H}_\mathrm{eff})\rho)-\sum_n\Tr(L_n \rho L_n^\dagger)}{\hbar}.
\end{eqnarray}
It follows
\begin{eqnarray}
\Tr\left( 2\Im(\hat{H}_\mathrm{eff})\rho- \sum_nL_n^\dagger L_n \rho\right)=0
\end{eqnarray}
or
\begin{eqnarray}
\Im(\hat{H}_\mathrm{eff})=\frac{1}{2}\sum_n L_n^\dagger L_n. \label{SM::Ln}
\end{eqnarray}
Substituting \eq{SM::Ln} into \eq{SM::Lindbladeq}, one gets a master equation of the conventional Lindblad form,
\begin{equation}
    i\hbar \frac{\partial \rho}{\partial t}=[\Re(\hat{H}_\mathrm{eff}),\rho]-\frac{i}{2}\sum_n 
    \left(L_n^{\dagger} L_n \rho + \rho L_n^{\dagger} L_n
    -2L_n \rho L_n^\dagger
    \right).
    \label{SM::Lindbladeq2}
\end{equation}
An explicit expression for $L_n$ can be read off by comparing \eq{SM::Ln} with \eq{SM::ImH}:
\begin{align}
    &\sum_n\rightarrow\int\mathrm{d}^3r\mathrm{d}^3r',\\
    &L_n\rightarrow L_{|\mathbf{r}-\mathbf{r'}|}=\sqrt{\Im(U(|\mathbf{r}-\mathbf{r'}|))}\Psi(\mathbf{r'})\Psi(\mathbf{r}).
\end{align}

In second quantization, the total number of particles $N$ is given by the number operator $\hat{N}$, 
\begin{equation}
    \hat{N}=\int\mathrm{d}^3r\Psi^\dagger(\mathbf{r})\Psi(\mathbf{r}).
\end{equation}
Its expectation value is given by $\expectation{\hat{N}}=\Tr(\rho \hat{N})$, where $\rho$ depends on the temperature $T$. Multiplying \eq{SM::Lindbladeq} by $\hat{N}$, subsequently taking the trace, and using that $\partial \hat{N}/\partial t$ is equal to zero in the Schr\"{o}dinger~ picture, we obtain 
\begin{equation}
    i\hbar \frac{\partial (\Tr(\rho \hat{N}))}{\partial t}=\Tr([\Re(\hat{H}_\mathrm{eff}),\rho]\hat{N})+i\Tr(\{\Im(\hat{H}_\mathrm{eff}),\rho\}\hat{N})-i\Tr \left(\sum_n L_n \rho L_n^\dagger \hat{N} \right).
    \label{SM::losseq1}
\end{equation}
 Note that the first term on the right hand side of \eq{SM::losseq1} is zero due to the commutativity between $\hat{N}$ and $\hat{H}_{\text{eff}}$:
\begin{equation}
    \begin{aligned}
    \mathrm{Tr}([\Re(\hat{H}_\mathrm{eff}),\rho]\hat{N})&=\mathrm{Tr}(\Re(\hat{H}_\mathrm{eff})\rho \hat{N})-\mathrm{Tr}(\rho \Re(\hat{H}_\mathrm{eff})\hat{N})=\mathrm{Tr}(\hat{N} \Re(\hat{H}_\mathrm{eff})\rho)-\mathrm{Tr}(\Re(\hat{H}_\mathrm{eff}) \hat{N}\rho)\\
    &=\mathrm{Tr}([\hat{N},\Re(\hat{H}_\mathrm{eff})]\rho)=0.
    \end{aligned}
\end{equation}
Thus, \eq{SM::losseq1} reduces to
\begin{equation}
    \hbar \frac{\partial \langle \hat{N} \rangle}{\partial t}=\mathrm{Tr}(\{\hat{N},\Im(\hat{H}_\mathrm{eff})\}\rho)-\mathrm{Tr}\left(\sum_n L_n^\dagger \hat{N} L_n \rho\right).
    \label{SM::losseq2}
\end{equation}
Rearranging the field operators in the expressions that appear on the right hand side of \eq{SM::losseq2} to be normal ordered,
\begin{equation}
    \begin{aligned}
    \{\hat{N},\Im(\hat{H}_\mathrm{eff})\}&=\int\mathrm{d}^3r\int\mathrm{d}^3r'\int\mathrm{d}^3r''\Im(U(|\mathbf{r}-\mathbf{r'}|))\Psi^\dagger(\mathbf{r})\Psi^\dagger(\mathbf{r'})\Psi^\dagger(\mathbf{r''})\Psi(\mathbf{r''})\Psi(\mathbf{r'})\Psi(\mathbf{r})\\
    &+2\int\mathrm{d}^3r\int\mathrm{d}^3r'\Im(U(|\mathbf{r}-\mathbf{r'}|))\Psi^\dagger(\mathbf{r})\Psi^\dagger(\mathbf{r'})\Psi(\mathbf{r'})\Psi(\mathbf{r}),
    \end{aligned}
\end{equation}
\begin{equation}
    \sum_nL_n^\dagger \hat{N} L_n= \int\mathrm{d}^3r\int\mathrm{d}^3r'\int\mathrm{d}^3r''\Im(U(|\mathbf{r}-\mathbf{r'}|))\Psi^\dagger(\mathbf{r})\Psi^\dagger(\mathbf{r'})\Psi^\dagger(\mathbf{r''})\Psi(\mathbf{r''})\Psi(\mathbf{r'})\Psi(\mathbf{r}),
\end{equation}
\eq{SM::losseq2} simplifies significantly:
\begin{equation}
    \hbar \frac{\partial \langle \hat{N} \rangle}{\partial t}=2\Tr\left(\int\mathrm{d}^3r\int\mathrm{d}^3r'\Im(U(|\mathbf{r}-\mathbf{r'}|))\Psi^\dagger(\mathbf{r})\Psi^\dagger(\mathbf{r'})\Psi(\mathbf{r'})\Psi(\mathbf{r})\rho\right).
    \label{SM::losseq}
\end{equation}
The right-hand side of \eq{SM::losseq} can be rewritten compactly in terms of the imaginary part of the effective Hamiltonian:
\begin{equation}
    \hbar \frac{\partial \langle \hat{N} \rangle}{\partial t}=
    4\expectation{\Im(\hat{H}_\mathrm{eff})};
    \end{equation}
While we mostly work with the homogeneous system, we note that external potentials can be readily added to Eq.~(\ref{Hamiltonian}) because only $U$ contributes to the imaginary part in the above equation.

Since we are interested in determining losses, our task is to derive an explicit expression for $\expectation{\Im(\hat{H}_\mathrm{eff})}=\Tr(\rho\Im(\hat{H}_\mathrm{eff}))$ for a weakly-interacting $p$-wave gas. Within the $p$-wave effective range theory, $U$ is a function of $v_p$ and $R$. Correspondingly, the operator $\hat{H}_\mathrm{eff}$ is also a function of $v_p$ and $R$: $\hat{H}_\mathrm{eff}=\hat{H}_\mathrm{eff}(v_p,R)$. When the absolute values of the dimensionless quantities $\mathcal{V}=n v_p$ and $\mathcal{R}= n^{1/3} R$ are both small, one can obtain
$\Im(\hat{H}_\mathrm{eff})$ by analytically continuing 
$\hat{H}_\mathrm{eff}(\mathcal{V},\mathcal{R})$ around $(\Re(\mathcal{V}),\Re(\mathcal{R}))$.
Treating $\hat{H}_{\text{eff}}$ as a function of the dimensionless quantities $\mathcal{V}$ and $\mathcal{R}$, the analytic continuation reads
\begin{equation}
    \Im(\hat{H}_\mathrm{eff}(\mathcal{V},\mathcal{R}))=\partialfix{\hat{H}_\mathrm{eff}(\Re(\mathcal{V}),\Re(\mathcal{R}))}{\mathcal{V}}{\mathcal{R}}\Im(\mathcal{V})+\partialfix{\hat{H}_\mathrm{eff}(\Re(\mathcal{V}),\Re(\mathcal{R}))}{\mathcal{R}}{\mathcal{V}}\Im(\mathcal{R})
    \label{SM::ana1}
    \end{equation}
    or
    \begin{equation}
    \Im(\hat{H}_\mathrm{eff}(\mathcal{V},\mathcal{R}))=\partialfix{\hat{H}_\mathrm{eff}(\Re(v_p),\Re(R))}{v_p}{R}\Im(v_p)+\partialfix{\hat{H}_\mathrm{eff}(\Re(v_p),\Re(R))}{R}{v_p}\Im(R).
    \label{SM::ana2}
\end{equation}
It is important to note that the arguments of $\hat{H}
_{\text{eff}}$ on the right hand sides of Eqs.~(\ref{SM::ana1})
and (\ref{SM::ana2}) are the {\em{real parts}} of the dimensionless quantities $\mathcal{V}$ and $\mathcal{R}$ and the {\em{real parts}} of the dimensionful scattering volume $v_p$ and effective range $R$, respectively.

Because we are concerned with the variation of the total number of particles at finite temperature, it is convenient to work in the grand canonical ensemble. The partition function $\mathcal{Z}$ and thermal state density matrix $\rho$ are given by~\cite{zagoskin1998quantum,fetter2012quantum}
\begin{align}
    &\mathcal{Z}=\exp\left(-\frac{\Omega}{k_B T}\right)=\Tr \left( \exp\left[\frac{-(\hat{H}_\mathrm{eff}-\mu\hat{N})}{k_B T}\right] \right),\label{SM::ZG}\\
    &\rho=\exp\left[\frac{\Omega-(\hat{H}_\mathrm{eff}-\mu\hat{N})}{k_B T}\right],\label{SM::rhoG}
\end{align}
where $\Omega$ and $\mu$ are the grand potential and chemical potential, respectively. In what follows, we will prove that 
\begin{equation}
    \expectation{\Im(\hat{H}_\mathrm{eff}(v_p,R))}=\partialfix{F(\Re(v_p),\Re(R))}{v_p}{R}\Im(v_p)+\partialfix{F(\Re(v_p),\Re(R))}{R}{v_p}\Im(R)
    \label{SM::expectationH}
\end{equation}
 holds in the weak interaction limit.
 In \eq{SM::expectationH}, $F=\Omega+\mu\expectation{\hat{N}}$ denotes the Helmholtz free energy. 
 \Eq{SM::expectationH} is important since it will subsequently allow us to relate the partial derivatives of the Helmholtz free energy to $p$-wave contacts.
 Treating---consistent with our effective $p$-wave theory framework---$\Omega$ and $\mu$ as functions of $v_p$ and $R$,
 we write
\begin{align}
    &\partialfix{F}{v_p}{R}=\partialfix{(\Omega+\mu\expectation{\hat{N}})}{v_p}{R}=\partialfix{\Omega}{v_p}{R}+\partialfix{\mu}{v_p}{R}\expectation{\hat{N}},\label{SM::F1}\\
    &\partialfix{F}{R}{v_p}=\partialfix{(\Omega+\mu\expectation{\hat{N}})}{R}{v_p}=\partialfix{\Omega}{R}{v_p}+\partialfix{\mu}{R}{v_p}\expectation{\hat{N}}.\label{SM::F2}
\end{align}
Applying the definitions from Eqs.~(\ref{SM::ZG}) and (\ref{SM::rhoG}), the partial derivative of the grand potential with respect to $v
_p$ while holding $R$ constant can be related to that of the effective Hamiltonian and the chemical potential with respect to the same variable, taken also while holding $R$ constant: 
\begin{equation}
    \begin{aligned}
    \partialfix{\Omega}{v_p}{R}&=-k_B T\partialfix{\ln\mathcal{Z}}{v_p}{R}\\
    &=-\frac{k_B T}{\mathcal{Z}}\frac{\partial}{\partial v_p}\left[\Tr \left(\exp\left(\frac{-\hat{H}_\mathrm{eff}+\mu\hat{N}}{k_B T}\right)\right)\right]_R\\
    &=\Tr\left[\exp\left(\frac{\Omega-\hat{H}_\mathrm{eff}+\mu\hat{N}}{k_B T}\right)\partialfix{\hat{H}_\mathrm{eff}}{v_p}{R}\right]\\
    &-\Tr\left[\exp\left(\frac{\Omega-\hat{H}_\mathrm{eff}+\mu\hat{N}}{k_B T}\right)\hat{N}\right]\partialfix{\mu}{v_p}{R}\\
    &=\left\langle\partialfix{\hat{H}_\mathrm{eff}}{v_p}{R}\right\rangle-\expectation{\hat{N}}\partialfix{\mu}{v_p}{R}.
    \end{aligned}
    \label{SM::F3}
\end{equation}
Similarly, one can find
\begin{equation}
    \partialfix{\Omega}{R}{v_p}=\left\langle\partialfix{\hat{H}_\mathrm{eff}}{R}{v_p}\right\rangle-\expectation{\hat{N}}\partialfix{\mu}{R}{v_p}.
    \label{SM::F4}
\end{equation}
Combining Eqs.~(\ref{SM::F1})-(\ref{SM::F4}) and (\ref{SM::ana2}), we obtain \eq{SM::expectationH}; this completes the proof of \eq{SM::expectationH}.

As already alluded to above, the next step is to relate the partial derivatives in \eq{SM::expectationH} that involve the Helmholtz free energy $F$ to $p$-wave contacts. The $p$-wave contacts are defined through the following equations~\cite{yu2015universal,yoshida2015universal,luciuk2016evidence}:
\begin{align}
    &\partialfix{F}{v_p^{-1}}{R}=-\frac{\hbar^2}{2m}\sum_m C_v^{(m)},\\
    &\partialfix{F}{R^{-1}}{v_p}=-\frac{\hbar^2}{2m}\sum_m C_R^{(m)},
\end{align}
where the superscript $m=-1,0,1$ represents the three $p$-wave scattering sub-channels. For a system with isotropic two-body interactions---as assumed throughout this work---, the sub-channel contacts are equal to each other: $C_v^{(m)}=C_v$ and $C_R^{(m)}=C_R$.
We thus have
\begin{align}
    &\partialfix{F}{v_p^{-1}}{R}=-\frac{3\hbar^2}{2m}C_v,\\
    &\partialfix{F}{R^{-1}}{v_p}=-\frac{3\hbar^2}{2m}C_R.
\end{align}
Changing the variables of the derivatives from $v_p^{-1}$ and $R^{-1}$ to $v_p$ and $R$, the partial derivatives of $F$ in \eq{SM::expectationH} can be expressed in terms of the $p$-wave contacts:
\begin{equation}
\begin{aligned}
    \frac{\mathrm{d}N}{\mathrm{d}t}&=\frac{4}{\hbar}\left(\partialfix{F}{v_p}{R}\Im(v_p)+\partialfix{F}{R}{v_p}\Im(R)\right)\\
    &=-\frac{4}{\hbar}\left(\frac{1}{(\Re(v_p))^2}\partialfix{F}{v_p^{-1}}{R}\Im(v_p)+\frac{1}{(\Re(R))^2}\partialfix{F}{R^{-1}}{v_p}\Im(R)\right)\\
    &=\frac{6\hbar}{m}\left(C_v\frac{\Im(v_p)}{(\Re(v_p))^2}+C_R\frac{\Im(R)}{(\Re(R))^2}\right).
\end{aligned}
\label{SM::contactloss}
\end{equation}

\section{Second-order virial expansion for homogeneous and trapped systems}\label{SMsect::virial}
This section is devoted to explaining the second-order virial expansion applied in this work. The virial expansion is applicable when the fugacity $z=\exp(\mu/k_B T)$ is small compared to $1$. This condition is equivalent to demanding that $n \lambda^3$, where $\lambda$ denotes the thermal wave length (see below for its definition), is small compared to $1$. This condition is fulfilled at high temperatures. Taylor-expanding the grand potential $\Omega$ up to second order in $z$, the virial expansion for the grand potential $\Omega$ and the expression for the second virial coefficient $b_2$ of the interacting system read~\cite{liu2013virial}
\begin{align}
    &\Omega=-k_B T Q_1\left(z+b_2z^2\right),\label{SM::virialomega}\\
    &b_2=\frac{Q_2-Q_1^2/2}{Q_1}.\label{SM::b2}
\end{align}
Here,  $Q_n$ denotes the canonical partition function of the $n$-body system.

\textit{Homogeneous system.---}
To calculate $b_2$, it is convenient to divide it into two terms, namely $b_2=b_2^{(0)}+\Delta b_2$, where  $b_2^{(0)}$ is the second virial coefficient of the non-interacting system 
($b_2^{(0)}$ accounts for the Fermi statistics) and $\Delta b_2$ encapsulates the effects of the two-body interactions. The quantity $b_2^{(0)}$ can be obtained from the exact expression of the grand potential $\Omega^{(0)}$ of the non-interacting single-component Fermi gas~\cite{fetter2012quantum}, 
\begin{equation}
    \Omega^{(0)}=-V \frac{k_B T}{\lambda^3}\frac{2}{\sqrt{\pi}}\int_0^\infty x^{1/2}\ln(1+z e^{-x})\mathrm{d}x. 
\end{equation}
Expanding the integrand and integrating term by term, $b_2^{(0)}$ can be shown to be equal to $\frac{-1}{2^{5/2}}$~\cite{liu2013virial}. 
The quantity $\Delta b_2$ can be written as $\Delta b_2=(Q_2-Q_2^{(0)})/Q_1$, where the superscript $(0)$ refers, again, to the non-interacting system. The one-body partition function $Q_1$, which appears in the denominator of the expression for $\Delta b_2$, can be straightforwardly calculated:
\begin{equation}
    Q_1=\frac{1}{h^3}\int \mathrm{d}^3r\mathrm{d}^3p \exp\left(-\frac{p^2}{2mk_B T}\right)=\frac{V}{\lambda^3},
    \label{SM::Q1}
\end{equation}
where $\lambda=\sqrt{2 \pi\hbar^2/(m k_B T)}$ 
is the thermal de Broglie wavelength. To evaluate $Q_2-Q_2^{(0)}$,  we use that the difference between the interacting and non-interacting partition functions is due to the relative motion (the center of mass motions are identical). This leads to the simplification
\begin{equation}
    Q_2-Q_2^{(0)}=\left[\frac{1}{h^3}\int \mathrm{d}^3 R\mathrm{d}^3 P\exp\left(-\frac{P^2}{4mk_B T}\right)\right]\left[\sum_{l,m} \int \mathrm{d}k \left(g_{l,m}(k)-g^{(0)}_{l,m}(k)\right)\exp\left({\frac{\hbar^2k^2}{m k_B T}}\right)\right],
    \label{SM::Q2Q20}
\end{equation}
where $\mathbf{R}$ and $\mathbf{P}$ denote the two-body center of
mass position and momentum vectors, respectively.
The first square bracket on the right hand side of \eq{SM::Q2Q20} is the partition function of the center-of-mass motion. It evaluates to $2^{3/2}V/\lambda^3$. The second square bracket on the right hand side of \eq{SM::Q2Q20} is the partition function of the relative motion. The quantities $g_{l,m}(k)$ and  $g^{(0)}_{l,m}(k)$ are the densities of states of the interacting and non-interacting two-body systems with relative orbital angular momentum $l$ and associated projection quantum number $m$. For polarized fermions, $l$ is restricted to odd values by the exchange symmetry. Since we focus on pure $p$-wave interactions in this work, $g_{l,m}(k)=g^{(0)}_{l,m}(k)$ for $l \geq 3$. Furthermore, the assumption of isotropic two-body interactions eliminates the dependence of $g_{l,m}(k)$ on $m$ and the summation over $m$ merely results in a multiplicative factor of $3$ for $l=1$. 

To calculate $g_{l=1}(k)-g^{(0)}_{l=1}(k)$, a hard sphere potential of radius $r_0$ is assumed. Such a potential imposes a boundary condition on the relative two-body wave function~\cite{huang2008statistical}. Using that the asymptotic $p$-wave scattering wave function reads~\cite{sakurai1995modern}
\begin{equation}
    \psi_{m}(\mathbf{r}) \xrightarrow{r\rightarrow\infty} A Y_{1,m}(\hat{\mathbf{r}})\frac{\cos(k r+\delta_p(k))}{r},
\end{equation}
the energy spectrum $\epsilon_n(k)=\hbar^2k_n^2/m$ 
of the hard sphere potential is determined by
\begin{equation}
    k_n r_0 +\delta_p(k_n)=\frac{\pi}{2}+n \pi, \mbox{ where } n=0,1,2,\cdots.
\end{equation}
Therefore, when $n$ increases by $1$, $k$ changes by $\Delta k=k_{n+1}-k_{n}$:
\begin{equation}
    \Delta k r_0+\delta_p(k_{n+1})-\delta_p(k_{n})=\pi.
\end{equation}
Dividing both sides by $\Delta k$ and taking the limit $r_0\rightarrow0$, one has $\Delta k\rightarrow dk$ and the above equation gives
\begin{equation}
    r_0+\dfrac{\mathrm{d}\delta_p(k)}{\mathrm{d}k}=\frac{\pi}{\mathrm{d}k}.
\end{equation}
Correspondingly, $\sum_n 1\rightarrow \int\mathrm{d}k g_{l=1}(k)$ and thus
\begin{equation}
    g_{l=1}(k)-g^{(0)}_{l=1}(k)=\frac{1}{\Delta k}-\frac{1}{\Delta k^{(0)}}=\frac{1}{\pi}\frac{\mathrm{d}\delta_p(k)}{\mathrm{d}k}.
    \label{SM::phaseshiftdiff}
\end{equation}
Combining Eqs.~(\ref{SM::b2}), (\ref{SM::Q1}), (\ref{SM::Q2Q20}) and Eq.~(\ref{SM::phaseshiftdiff}), we obtain the Beth-Uhlenbeck formula
\begin{equation}
    b_2=b_2^{(0)}+\frac{3\times2^{3/2}}{\pi}\int_0^\infty\mathrm{d}k \exp\left(-\frac{\hbar^2k^2}{mk_B T}\right)\frac{\mathrm{d}\delta_p(k)}{\mathrm{d}k}.
    \label{SM::BUformula}
\end{equation}
To perform the integral in Eq.~(\ref{SM::BUformula}), we expand ${\mathrm{d}\delta_p(k)}/{\mathrm{d}k}$ with the help of \eq{SM::phaseshift},
\begin{equation}
    \frac{\mathrm{d}\delta_p(k)}{\mathrm{d}k}=\frac{\mathrm{d}\delta_p(k)}{\mathrm{d}\tan(\delta_p(k))}\frac{\mathrm{d}\tan(\delta_p(k))}{\mathrm{d}k}=-3 v_p k^2 + \frac{5 v_p^2}{R}k^4.
    \label{SM::dddk}
\end{equation}
Substituting \eq{SM::dddk} into Eq.~(\ref{SM::BUformula}), $b_2$ is expressed in terms of $v_p$ and $R$,
\begin{equation}
    b_2=-\frac{1}{4\sqrt{2}}+\frac{18 \pi v_p (5 \pi v_p-\lambda^2 R)}{\lambda^5 R}.
    \label{SM::b2vR}
\end{equation}
Together with Eqs.~(\ref{SM::virialomega}) and (\ref{SM::Q1}), \eq{SM::b2vR} provides an explicit expression for the grand potential $\Omega$. To determine explicit expressions for the $p$-wave contacts, the grand potential needs to be reexpressed in terms of the Helmholtz free energy. To accomplish this task, the fugacity $z$ needs to be expressed in terms of $n=N/V$. By the Gibbs-Duhem relation~\cite{landau2013quantum}, one has
\begin{equation}
    N=-\frac{z}{k_B T}\partialfix{\Omega}{z}{{V,T}}.
    \label{SM::GDrelation}
\end{equation}
Using $\Omega=-PV$, where $P$ denotes the pressure, \eq{SM::GDrelation} can be rewritten as
\begin{equation}
    n=\frac{z}{k_B T}\partialfix{P}{z}{{T}}.
\end{equation}
Using the expansion of $\Omega$ in terms of $z$, we find the following expansions for the pressure and density:
\begin{equation}
    P=\frac{k_B T z(1+b_2 z)}{\lambda^3}
\end{equation}
and
\begin{equation}
    n=\dfrac{z}{\lambda^3}+2 b_2 \dfrac{z^2}{\lambda^3}.
    \label{SM::nzrelation}
\end{equation}
At this order of the expansion, an explicit expression for $z$ can be obtained from the algebraic equation \eq{SM::nzrelation}. The physical solution reads
\begin{equation}
    z=\dfrac{-1}{4 b_2}+\dfrac{\sqrt{1+8 b_2 \lambda^3 n }}{4 b_2}.
    \label{SM::znrelation}
\end{equation}
The free energy $F$ can then be calculated using $F=\Omega+\mu N=\Omega+n V k_B T \ln(z)$ and using the right hand side of \eq{SM::znrelation} to eliminate $z$ from
$\Omega+n V k_B T \ln(z)$; as a result, $F$ is expressed in terms of $n$. With this in hand, expressions for the two $p$-wave contacts  can be derived in terms of $n$, $\Re{(v_p)}$, and $\Re{(R)}$. To this end, we Taylor-expand the contacts---consistent with the virial expansion framework applied to the thermodynamic quantities---up to second order in $n\lambda^3$:
\begin{align}
    C_v&=\dfrac{2m(\Re(v_p))^2}{3\hbar^2}\partialfix{F}{v_p}{R}=\dfrac{12\pi m k_B T n^2 V (\Re(v_p))^2}{\hbar^2}+\dfrac{60\pi m^2 k_B^2 T^2 n^2 V (\Re(v_p))^3}{\Re(R) \hbar^4},\label{SM::virialCv}\\
    C_R&=\dfrac{2m(\Re(R))^2}{3\hbar^2}\partialfix{F}{R}{v_p}=\dfrac{30\pi m^2 k_B^2 T^2 n^2 V (\Re(v_p))^2}{\hbar^4}.\label{SM::virialCR}
\end{align}
We emphasize that the expressions for the two $p$-wave contacts in Eqs.~(\ref{SM::virialCv}) and (\ref{SM::virialCR}) apply to any $p$-wave system that is characterized by $v_p$ and $R$, including systems with arbitrarily large $|\Re(v_p)|$ and $|\Re(R)|$. Specifically, the assumption 
 of  weak interactions has not yet entered into the derivation. Importantly, in addition to the expected linear dependence of $C_v$ on $T$, $C_v$ and $C_R$ each contain a term that depends quadratically on $T$. Both these quadratic terms arise from the effective range: The $T^2$ term in $C_v$ explicitly depends on the effective range while the $T^2$ in $C_R$ arises from taking the derivative of $F$ with respect to $R$.
 Equations~(\ref{SM::virialCv}) and (\ref{SM::virialCR}) are identical to the expressions reported in Ref.~\cite{he2020universal}; in that work, it was suspected that the two $T^2$ terms may account for the suppression of $\beta/T$. In this context, though, it is important to keep in mind that the expressions for $C_v$ and $C_R$, Eqs.~(\ref{SM::virialCv}) and (\ref{SM::virialCR}), are derived within the high-temperature virial expansion approach while the suppression was experimentally observed in the low-temperature regime~\cite{demarco2019degenerate}. 
 
 We now consider the weakly-interacting regime by taking the limit of \eq{SM::inequality}. 
 Technically, we first express $\Re(R)$ to be $c_1(\Re(v_p))^{1/3}$, where $c_1$ is a constant of order one, and then evaluate Eqs.~(\ref{SM::virialCv}) and (\ref{SM::virialCR}) in the regime where $n \Re(v_p)\ll 1$. The results are 
\begin{align}
    C_v&=\dfrac{12\pi m k_B T n^2 V (\Re(v_p))^2}{\hbar^2},\label{SM::virialCv2}\\
    C_R&=\dfrac{30\pi m^2 k_B^2 T^2 n^2 V (\Re(v_p))^2}{\hbar^4}.\label{SM::virialCR2}
\end{align}
Substituting Eqs.~(\ref{SM::virialCv2}) and (\ref{SM::virialCR2}) into \eq{SM::contactloss}, $\mathrm{d}N/\mathrm{d}t$ and subsequently $\beta$ can be found. 
The result can be further simplified by repeating the progress above for the imaginary parts: expressing $\Im(R)$ as $c_2 (\Im(v_p))^{1/3}$, where $c_2$ is a real number, and taking the limit $n \Im(v_p)\ll1$, the term that includes $R$ disappears.

\textit{Harmonically trapped system.---} As demonstrated by \eq{SM::b2}, $b_2$ is determined by $Q_n$ with $n=1$ and $2$. This implies that one can get $b_2$ by solving the one- and two-body problems. \citet{busch1998two} reported the eigen states and eigen energies for two $s$-wave particles with zero-range pseudo-potential~\cite{huang1957quantummechanical} confined in a harmonic trap. The approach was subsequently generalized to two spin-polarized fermions in a harmonic trap interacting through a $p$-wave pseudo-potential~\cite{kanjilal2004nondivergent}. The energy spectrum for the relative motion was shown to be implicitly given by
\begin{equation}
    \frac{v_p}{a_{\mathrm{ho}}^3} 
    \left(1 - \cfrac{2 m v_p E_\mathrm{rel}}{\hbar^2 R} \right)=-\cfrac{\Gamma\left(-\dfrac{E_\mathrm{rel}}{2 \hbar \omega}-\cfrac{1}{4}\right)}{2\sqrt{2} \Gamma\left(-\cfrac{E_\mathrm{rel}}{2 \hbar \omega}+\cfrac{5}{4}\right)},
    \label{SM::pwavespectrum}
\end{equation}
where $\omega$ and $a_\mathrm{ho}=\sqrt{\hbar/(m \omega)}$ are the angular frequency of the harmonic trap and the length scale determined by it. In writing \eq{SM::pwavespectrum}, the phase shift is expressed using the effective range expansion. Similar to the discussion above, when $v_p\rightarrow 0$, the second term in round brackets on the left hand side of \eq{SM::pwavespectrum} scales as  $|v_p/R| a_\mathrm{ho}^2 \propto |v_p^{2/3}|/a_\mathrm{ho}^2< |v_p^{2/3}|n^{2/3} \ll 1$.  Dropping the energy-dependent term on the left hand side, \eq{SM::pwavespectrum} reduces to
\begin{equation}
    \cfrac{v_p}{a_\mathrm{h o}^3}=-\cfrac{\Gamma\left(-\dfrac{E_\mathrm{rel}}{2 \hbar \omega}-\cfrac{1}{4}\right)}{2\sqrt{2} \Gamma\left(-\cfrac{E_\mathrm{rel}}{2 \hbar \omega}+\cfrac{5}{4}\right)}.
    \label{SM::pwavespectrum2}
\end{equation}
When $v_p$ vanishes identically, \eq{SM::pwavespectrum2} recovers the $p$-wave energy spectrum of the non-interacting system~\cite{kanjilal2004nondivergent}:
\begin{equation}
    E_\mathrm{rel}^{(0)}(n,l=1)=\left(2n+\dfrac{5}{2}\right)\hbar\omega \qquad n=0,1,2, \cdots .
    \label{SM::spectrum0}
\end{equation}
The energy spectrum of the weakly-interacting system can thus be obtained by Taylor-expanding \eq{SM::pwavespectrum2} in $v_p/a_{\text{ho}}^3$ around the non-interacting relative energies,
\begin{equation}
    E_\mathrm{rel}(n,1)=E_\mathrm{rel}^{(0)}(n,1)+f(n)\frac{v_p}{a_{\text{ho}}^3},
\end{equation}
where $f(n)$ is given by $a_{\text{ho}}^3 \mathrm{d}E_\mathrm{rel}/\mathrm{d}v_p$, evaluated at $E_{\text{rel}}^{(0)}(n,1)$ with $n=0,1,2, \cdots$,
\begin{equation}
    f(n)=\lim_{E_\mathrm{rel} \rightarrow (5/2+2n)\hbar \omega} a_{\text{ho}}^3 \left(\frac{\mathrm{d}v_p}{\mathrm{d}E_\mathrm{rel}}\right)^{-1}=\frac{4\sqrt{2}}{(-1)^n n! \Gamma\left(-\cfrac{3}{2}-n\right)} \hbar \omega, \mbox{ where } n=0,1,2,\cdots .
\end{equation}
The two-body partition function $Q_2$ is then
\begin{equation}
    Q_2=Q_1 \left\{ \sum_{n=0}^\infty\left[ 3 \exp\left(-\dfrac{E_\mathrm{rel}(n,1)}{k_B T}\right)+\sum_{l=3,5,7...}(2l+1)\exp\left(-\dfrac{E^{(0)}_\mathrm{rel}(n,l)}{k_B T}\right)\right] \right\},
    \label{SM::trapQ2}
\end{equation}
where $E^{(0)}_\mathrm{rel}(n,l)=(2n+l+3/2)\hbar\omega$ is the non-interacting energy spectrum for the $l$th partial wave channel. The $Q_1$ in \eq{SM::trapQ2} is the partition function of the center-of-mass motion. Since the center-of-mass motion is identical to that of a non-interacting particle, its partition function is
equal to the one-body partition function:
\begin{equation}
    Q_1 = \sum_{n=0}^\infty\sum_{l=0}^\infty(2l+1)\exp\left(-\dfrac{E^{(0)}_\mathrm{rel}(n,l)}{k_B T}\right)=\frac{\exp\left({\dfrac{3\hbar\omega}{2k_B T}}\right)}{\left(\exp\left({\dfrac{\hbar\omega}{k_B T}}\right)-1\right)^3}.
\end{equation}
The expression inside the curly bracket on the right hand side of \eq{SM::trapQ2} represents the partition function of the relative motion. The summation only goes over odd relative orbital angular momentum quantum numbers $l$, since the spatial wave function of two polarized fermions (two fermions in the same spin state) must be anti-symmetric under the exchange of the two particles. The  sum $\sum_{l=3,5,\cdots}$ comes from the higher angular momentum states, which are by assumption not impacted by the interactions. The sum has a compact analytical expression:   
\begin{equation}
    \sum_{l=3,5,7...}(2l+1)\exp\left(-\dfrac{E^{(0)}_\mathrm{rel}(n,l)}{k_B T}\right)=\exp\left({\dfrac{-\hbar\omega}{2k_B T}}\right)\frac{\left[-3+7\exp\left({\dfrac{2\hbar\omega}{k_B T}}\right)\right]}{\left[\exp\left({\dfrac{2\hbar\omega}{k_B T}}\right)-1\right]^3}.
\end{equation}
 The first term in the curly brackets on the right hand side of \eq{SM::trapQ2} corresponds to $p$-wave states (the factor of $3$ is due due to the three projection quantum numbers $m=-1,0,1$). The sum is evaluated numerically by choosing a sufficiently large energy cutoff (i.e., maximal $n$ value). 

\begin{figure}
    \centering
    \includegraphics[width=0.5\textwidth]{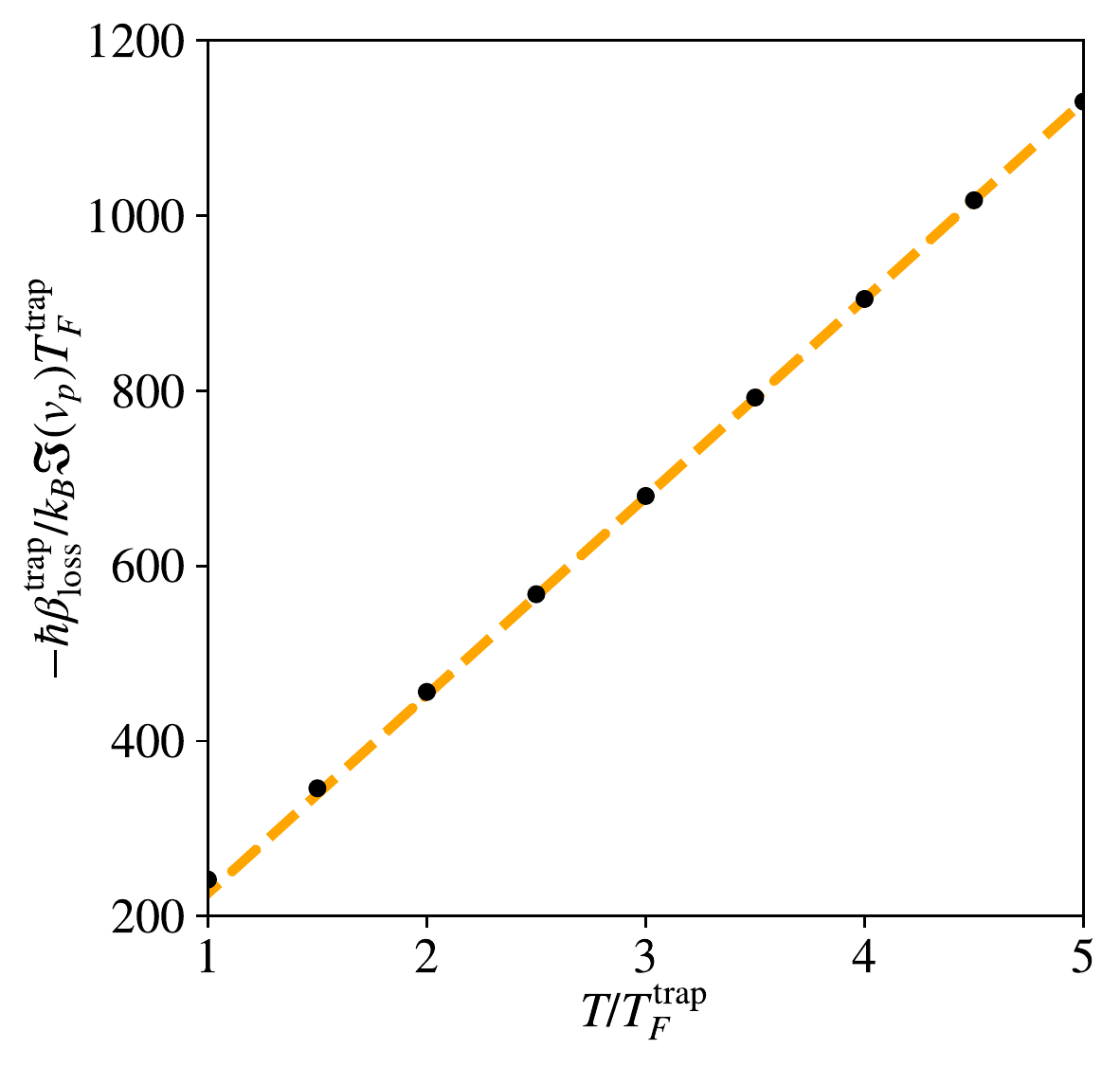}
    \caption{$\beta_\text{loss}^\text{trap}$  against $T$ in the high-$T$ regime.The dashed line and the black circles show Eq.~(\ref{betaTr}) from the main text and the prediction from the second-order virial expansion based on the exact two-body energy spectrum~\cite{kanjilal2004nondivergent}.}
    \label{SM::trapvirial}
\end{figure}
 
With $b_2$ evaluated, the subsequent steps proceed analogously to those for the homogeneous system. Specifically, $z$ needs to be converted to $N$ by solving \eq{SM::GDrelation}. The physical solution is
 \begin{equation}
     z=\frac{Q_1-\sqrt{Q_1^2-4NQ_1^2+8NQ_2}}{2(Q_1^2-2Q_2)}.
     \label{SM::ztrap}
 \end{equation}
Substituting \eq{SM::ztrap} into \eq{SM::virialomega}, we get
\begin{equation}
    \Omega=k_B T\dfrac{4NQ_2+Q_1\left[-(2N+1)Q_1+\sqrt{Q_1^2-4NQ_1^2+8NQ_2}\right]}{4(Q_1^2-2Q_2)}.
\end{equation}
Using $F=\Omega+\mu N$, $C_v$ is given by
\begin{equation}
         C_v=\frac{2m}{3\hbar^2}\frac{\partial F(\Re(v_p))}{\partial v_p}(\Re(v_p))^2=-\frac{m k_B T (\Re(v_p))^2}{3\hbar^2}\frac{4NQ_2-Q_1\left((2N-1)Q_1+\sqrt{Q_1^2-4NQ_1^2+8NQ_2}\right)}{(Q_1^2-2Q_2)^2}\frac{\partial Q_2(\Re(v_p))}{\partial v_p}.
\end{equation}
It follows that $\mathrm{d}N/\mathrm{d}t$ for the harmonically trapped system can be evaluated using Eq.~(\ref{contacteq}). This allows one to calculate $\beta_{\text{loss}}^{\text{trap}}$ [Eq.~(\ref{betaTr}) in the main text] directly without resorting to the local density approximation(LDA). \Fig{SM::trapvirial} shows excellent agreement between the black dots (virial expansion for the trapped system) and the dashed line [Eq.~(\ref{betaTr}) in the main text]. The result shown in \fig{SM::trapvirial} uses an energy cutoff of $4,000\hbar\omega$. The agreement provides an important validation of LDA framework applied in the main text.

\section{Landau Fermi liquid theory}
Our derivation builds on a recent work, which developed a description of the weakly-interacting homogeneous polarized Fermi gas using Fermi liquid theory~\cite{ding2019fermiliquid}. 
Here, their formalism is directly extended to the complex domain because, under the condition \eq{SM::inequality}, the second-order expansion of all thermodynamic quantities and Fermi liquid parameters are automatically analytic.
The ground state energy $E_0$ of the homogeneous $p$-wave gas is given by
\begin{equation}
    E_0=N E_F \bkt{\frac{3}{5}+\frac{6 k_F^3 v_p}{5 \pi}-\bkt{\frac{18}{35\pi k_F R}-\frac{2066-312\ln(2)}{1155\pi^2}}k_F^6v_p^2},
    \label{SM::E0}
\end{equation}
where $E_F=\hbar^2 k_F^2/2m $ and $k_F$ denote the Fermi energy and Fermi momentum, respectively, of the non-interacting system. 
To obtain the temperature dependence of the internal energy, one can integrate the heat capacity $c_V(T)$,
\begin{equation}
    U_\text{int}=E_0+V\int_0^{T}c_V(T')\mathrm{d}T'=E_0+\frac{\pi^2}{12}\nu(0) V k_B^2T^2,
    \label{SM::ET}
\end{equation}
where~\cite{baym2008landau}
\begin{equation}
    c_V(T)=\frac{\pi^2}{6} \nu(0) k_B^2 T.
    \label{SM::cVT}
\end{equation}
Here, $\nu(0)$ is the density of states at the Fermi surface, which is determined by the effective mass $m^{*}$ and the Fermi wave vector $k_F$, 
\begin{equation}
    \nu(0)=\frac{m^*k_F}{2\pi^2\hbar^2}.
\end{equation}
The effective mass reads~\cite{ding2019fermiliquid}
\begin{equation}
    \frac{m}{m^*}=1+\frac{2 k_F^3}{\pi} v_p+\bkt{\frac{2}{\pi k_F R}-\frac{8(313-426\ln(2)}{315\pi^2}}k_F^6v_p^2.
    \label{SM::mstar}
\end{equation}
Since our aim is to derive expressions for the $p$-wave contacts by taking partial derivatives of the Helmholtz free energy, we need to ``switch" from ``energy" to ``Helmholtz free energy." This can be accomplished via the relation $F=U_\text{int}-TS$, where $S$ denotes the entropy. We start with the definition of the entropy 
\begin{equation}
\begin{split}
    S&=-k_B \sum_{\mathbf{p}} \left[ n_{\mathbf{p}}\ln{n_{\mathbf{p}}}+(1-n_{\mathbf{p}})\ln{(1-n_{\mathbf{p}})} \right]\\
    &=\frac{k_B V}{2\pi^2\hbar^3}\int_0^{+\infty}\mathrm{d}p p^2 n_p \ln \left( {\frac{1-n_p}{n_p}} \right)-\frac{k_B V}{2\pi^2\hbar^3}\int_0^{+\infty}\mathrm{d}p p^2 \ln{(1-n_p)},\\
\end{split}
\end{equation}
where $n_{\mathbf{p}}$ is the quasi-particle momentum distribution function. In the last equal sign of the last equation, we assumed a spherically symmetric quasi-particle momentum distribution function and converted the sum over $\mathbf{p}$  to an integral. Changing the variable $p$ to the energy $\epsilon_p$ using the relation $\mathrm{d} \epsilon_p = (p/m^*) \mathrm{d} p$, we find
\begin{equation}
    S=\frac{k_B V}{2\pi^2\hbar^3}\int_0^{+\infty}\mathrm{d}\epsilon_p m^* p n_p \ln \left( {\frac{1-n_p}{n_p}} \right)-\frac{k_B V}{2\pi^2\hbar^3}\int_0^{+\infty}\mathrm{d}\epsilon_p m^* p \ln{(1-n_p)}.
\end{equation}
Applying the Sommerfeld expansion to the first term of the above expression, we find
\begin{equation}
    S=\frac{k_B V}{2\pi^2\hbar^3}\int_0^{\mu}\mathrm{d}\epsilon_p m^* p \ln \left( {\frac{1-n_p}{n_p}} \right)+\frac{V m^* p_F}{12\hbar
    ^3}k_B^2 T-\frac{k_B V}{2\pi^2\hbar^3}\int_0^{+\infty}\mathrm{d}\epsilon_p m^* p \ln{(1-n_p)}.
\end{equation}
In the weakly-interacting limit ($n|v_p|\ll 1$), the quasi-particle residue would approach zero. Thus, as $T \ll T_F$, we have $n_p\simeq1$ for $p<p_F$ and $n_p\simeq0$ for $p>p_F$. Hence, the first term and the third term above can be canceled to get
\begin{equation}
\begin{split}
    S&=\frac{V m^* p_F}{12\hbar^3}k_B^2 T\\
    &=\frac{\pi^2}{6}\nu(0)Vk_B^2 T.\\
\end{split}
\end{equation}
Inserting the above result for $S$, the Helmholtz free energy of the system becomes
\begin{equation}
    F=U_\text{int}-TS=E_0-\frac{\pi^2}{12}\nu(0)Vk_B^2T^2.
\end{equation}
The two $p$-wave contacts can now be obtained by taking partial derivatives of the free energy. We find
\begin{align}
    C_v&=\dfrac{12\times6^{2/3}\pi^{7/3} n^{8/3} V (\Re(v_p))^2}{5}+\dfrac{2^{1/3}\pi^{5/3}m^2k_B^2T^2n^{4/3} V (\Re(v_p))^2}{3^{2/3}\hbar^4},\label{SM::CvFermi}\\
    C_R&=\dfrac{216\times6^{1/3}\pi^{11/3} n^{10/3} V (\Re(v_p))^2}{35}+\dfrac{2\pi^3m^2k_B^2T^2n^2 V (\Re(v_p))^2}{\hbar^4}.\label{SM::CRFermi}
\end{align}
Note that both Eqs.~(\ref{SM::CvFermi}) and (\ref{SM::CRFermi}) are calculated up to the second order in $n\Re(v_p)$, after restricting $\Re(R)\propto (\Re(v_p))^{1/3}$. Similar to the case at high temperatures, the contribution to $\mathrm{d}N/\mathrm{d}t$ and $\beta$ from $C_R$ can be discarded by taking $n\Im(v_p)<<1$ after substituting Eqs.~(\ref{SM::CvFermi}) and (\ref{SM::CRFermi}) into \eq{SM::contactloss}.

It is worth noting that the $T^2$ terms from Eqs.~(\ref{SM::CvFermi}) propagate into the expression for $\beta$; they do not---as in \eq{SM::virialCv2}, which is applicable to the weakly interacting regime,---become negligible. 
In this sense, Ref.~\cite{he2020universal} indeed gives the correct insight that a $T^2$ term might play a role in $C_v$. We emphasize, however, that the mechanisms behind the $T^2$ terms in Eqs.~(\ref{SM::CvFermi}) and (\ref{SM::virialCv2}) are completely different than those considered in Ref.\cite{he2020universal}.

\section{Local-density approximation}
\label{SMsect::LDA}

The phase space density of the non-interacting homogeneous single-component Fermi gas follows the Fermi-Dirac distribution
\begin{equation}
    w(\mathbf{r},\mathbf{p})=\frac{1}{(2\pi)^3}\dfrac{1}{\exp[(p^2/2m-\mu)/k_B T]+1}.
\end{equation}
Using this, the phase space density of the inhomogeneous system (isotropic harmonic trap with angular trap frequency $\omega$) can be obtained within the Thomas-Fermi approximation or LDA~\cite{butts1997trapped}:
\begin{equation}
    \begin{split}
        w(\mathbf{r},\mathbf{p})&=\frac{1}{(2\pi)^3}\dfrac{1}{\exp[(p^2/2m-\mu(\mathbf{r}))/k_B T]+1}\\
        &=\frac{1}{(2\pi)^3}\dfrac{1}{\exp[(p^2/2m+m\omega^2r^2/2-\mu)/k_B T]+1}.
    \end{split}
\end{equation}
The real space density is obtained by integrating out the momentum dependence:
\begin{equation}
    \begin{split}
    n(\mathbf{r})&=\frac{1}{(2\pi\hbar)^3}\int\mathrm{d}^3p\dfrac{1}{\exp[(p^2/2m+m\omega^2r^2/2-\mu)/k_B T]+1}\\
    &=-\dfrac{(m k_B T)^{3/2}\mathrm{Li}_{3/2}\left[-\exp \left( \dfrac{2\mu-m\omega^2 r^2}{2k_B T}\right)\right]}{(2\pi)^{3/2}\hbar^3},
    \end{split}
    \label{SM::nr}
\end{equation}
where $\mathrm{Li}$ denotes the polylog function. The chemical potential $\mu$ is determined by the restriction on the total number of particles $N$,
\begin{equation}
    N=\int \mathrm{d}\epsilon \dfrac{g(\epsilon)}{\exp[(\epsilon-\mu)/k_B T]+1},
\end{equation}
where $g(\epsilon)=\epsilon^2/[2(\hbar\omega)^3]$ is the density of states~\cite{butts1997trapped}. The explicit form of $\mu$ is 
\begin{equation}
    \mu=k_B T\ln\left[-\mathrm{Li}_{3}^{-1}\left(-\dfrac{(\hbar\omega)^3N}{(k_B T)^3}\right)\right].
    \label{SM::mu}
\end{equation}
$\mathrm{Li}_{3}^{-1}$ represents the inverse function of $\mathrm{Li}_3$, i.e. $y=f^{-1}(x)$ indicates $x=f(y)$.

In practice, the $T\simeq 0$ case needs special attention due to the difficulty of evaluating \eq{SM::nr} numerically. In this case, one can start from the chemical potential, which is---at $T=0$---equal to the Fermi energy:
\begin{equation}
    \mu = E_F = (6N)^{1/3}\hbar\omega.
\end{equation}
In the homogeneous system, the relationship between the chemical potential of the ground state and the density is
\begin{equation}
    n=\frac{1}{6\pi^2}\left(\dfrac{2m\mu}{\hbar^2}\right)^{3/2}.
\end{equation}
Using the LDA, the density profile of the trapped system at $T=0$ is
\begin{equation}
    n(\mathbf{r})= \frac{1}{6\pi^2}\left(\dfrac{48^{1/3}N^{1/3}a_{\mathrm{ho}}^2-r^2}{a_{\mathrm{ho}}^4}\right)^{3/2} \mbox{ for } r<R_F=(48N)^{1/6}a_{\mathrm{ho}}.
\end{equation}
The cutoff or Thomas-Fermi radius $R_F$ is determined by the condition $n(\mathbf{r}) \ge 0$.

With the expressions for $n(\mathbf{r})$ given above
(either the $T>0$ or the $T=0$ expression), $\beta_\text{loss}^\text{trap}$ can be calculated using Eq.~(\ref{betaTr}) from the main text.

Interpreting $\ntrap$ to be a functional of $n(\mathbf{r})$, $\betaT$ can be calculated directly within the LDA. We start with the definition of $\ntrap$ [see Table~(\ref{SM::notationlist})]:
\begin{equation}
    \ntrap[n(\mathbf{r})]=\dfrac{\int\mathrm{d}^3r[n(\mathbf{r})]^2}{\int\mathrm{d}^3rn(\mathbf{r})}.
\end{equation}
The functional derivative and functional differential of $\ntrap$ are
\begin{align}
    &\dfrac{\delta \ntrap}{\delta n(\mathbf{r})}=\dfrac{2 n(\mathbf{r})}{\int\mathrm{d}^3rn(\mathbf{r})}-\dfrac{\int\mathrm{d}^3r[n(\mathbf{r})]^2}{\left(\int\mathrm{d}^3rn(\mathbf{r})\right)^2},\label{SM::funcderi}\\
    &\mathrm{d}\ntrap=\int\mathrm{d}^3r\dfrac{\delta \ntrap}{\delta n(\mathbf{r})}\delta n(\mathbf{r}).\label{SM::funcdiff}
\end{align}
The variation of $n(\mathbf{r})$ is then governed by the homogeneous two-body loss coefficient $\beta(\mathbf{r})$, which is defined by
 \begin{equation}
     \delta n(\mathbf{r})=-\beta(\mathbf{r}) [n(\mathbf{r})]^2 \mathrm{d}t,\label{SM::localloss}
 \end{equation}
 where $\beta(\mathbf{r})$ is interpreted to be the \textit{homogeneous} $\beta$ evaluated with a local Fermi temperature 
 \begin{equation}
     T_F(\mathbf{r})=\frac{\hbar^2}{2m k_B} [6\pi^2\nT]^{2/3}.
     \label{SM::localTF}
 \end{equation}
 For any $\mathbf{r}$, with Eqs.~(\ref{SM::nr}), (\ref{SM::localTF}) and the result reported in Fig.~1 of the main text, $\beta(\mathbf{r})$ can be evaluated. Substituting \eq{SM::funcdiff} into the definition of $\betaT$ [see Table~(\ref{SM::notationlist})], we obtain
 \begin{equation}
    \betaT=\dfrac{2\left(\int\mathrm{d}^3rn(\mathbf{r})\right)\left(\int\mathrm{d}^3r\beta(\mathbf{r})[n(\mathbf{r})]^3\right)}{\left(\int\mathrm{d}^3r[n(\mathbf{r})]^2\right)^2}-\dfrac{\int\mathrm{d}^3r\beta(\mathbf{r})[n(\mathbf{r})]^2}{\int\mathrm{d}^3r[n(\mathbf{r})]^2}.
    \label{SM::betaT}
 \end{equation}
 Comparing with the definitions of $\betaT$ and $\beta_\text{loss}^\text{trap}$, the excessive part that is due to the change of the volume is identified as
 \begin{equation}
     \beta_\mathrm{deform}^\mathrm{trap}= \dfrac{2\left(\int\mathrm{d}^3rn(\mathbf{r})\right)\left(\int\mathrm{d}^3r\beta(\mathbf{r})[n(\mathbf{r})]^3\right)}{\left(\int\mathrm{d}^3r[n(\mathbf{r})]^2\right)^2}-\dfrac{2\int\mathrm{d}^3r\beta(\mathbf{r})[n(\mathbf{r})]^2}{\int\mathrm{d}^3r[n(\mathbf{r})]^2}.
     \label{SM::betaTV}
 \end{equation}
When $\beta(\mathbf{r})$ becomes independent of $\mathbf{r}$, which is the case in the high-temperature regime, $\betaT$ and $\beta_\mathrm{deform}^\mathrm{trap}$ are proportional to the loss coefficient $\beta$ of the homogeneous system. At high temperature, \eq{SM::nr} reduces to
\begin{equation}
    n(\mathbf{r})=\dfrac{m^{3/2}\omega^3 N\exp\left(\dfrac{-m\omega^2 r^2}{2k_B T}\right)}{(2\pi k_B T)^{3/2}}.
    \label{SM::nrcls}
\end{equation}
In this case, the integrals in \eq{SM::betaT} and \eq{SM::betaTV} can be explicitly performed. The results are
\begin{align}
    & \betaT= \left(\dfrac{16}{3\sqrt{3}}-1\right)\beta,\\
    & \beta_\text{deform}^\text{trap}=\left(\dfrac{16}{3\sqrt{3}}-2\right)\beta.
\end{align}


There is an illuminating way of writing Eqs.~(\ref{SM::betaT}) and (\ref{SM::betaTV}) at high temperatures. Using $\langle\dots\rangle$ to denote an average over the density profile [$\langle\dots\rangle=N^{-1}\int d^3r \dots n(\mathbf r)$], the equations can be rearranged into
\begin{align}
    \beta^\text{trap}&\xrightarrow{T/T_F\gg1}\beta\left(\dfrac{2\langle n^2(\mathbf r)\rangle-\langle n(\mathbf r)\rangle^2}{\langle n(\mathbf r)\rangle^2}\right),\\
    \beta^\text{trap}_\text{deform}&\xrightarrow{T/T_F\gg1}2\beta\left(\dfrac{\langle n^2(\mathbf r)\rangle-\langle n(\mathbf r)\rangle^2}{\langle n(\mathbf r)\rangle^2}\right).\label{SM::highTbetaTd}
\end{align}
Equation~(\ref{SM::highTbetaTd}) indicates that $\beta^\text{trap}_\text{deform}$ is proportional to the variance of the local density of the system at high temperatures. It should be emphasized that the variance of the local density is not equivalent to the ``density fluctuations" discussed in Ref.~\cite{demarco2019degenerate}; the latter are particle fluctuations \textit{at fixed chemical potential} of the homogeneous system. In this work, as we used the Helmholtz free energy to define the contact within the canonical ensemble, the number of particles is fixed. Therefore, our picture is different from the conjecture that density fluctuations suppress the loss proposed in Ref.~\cite{demarco2019degenerate}. 

\begin{figure}
    \centering
    \includegraphics[width=0.5\textwidth]{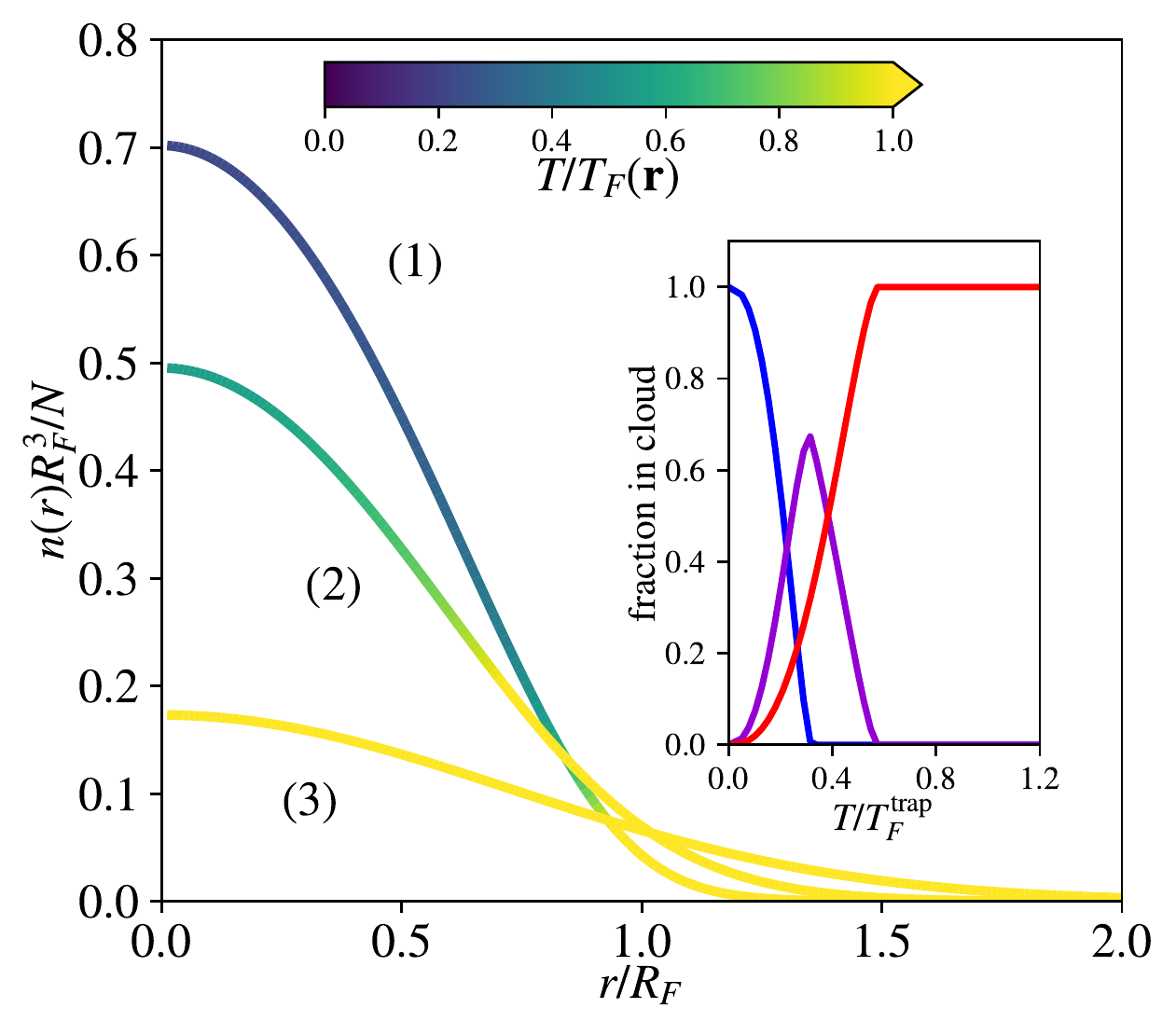}
    \caption{Density profiles for $T/T_F^\mathrm{trap}=0.2$, $0.4$, and $1$
    [curves labeled by (1), (2), and (3), respectively; $R_F=(48N)^{1/6}{\sqrt{\hbar/m \bar\omega}}$ is the radius of the $T=0$ cloud]. Inset: The blue, purple, and red lines show the fraction of the inhomogeneous loss-rate coefficient due to the high-, medium-, and low-temperature theory of the homogeneous system as a function of the local temperature.}
    \label{SM::local_TF}
\end{figure}

The lines labeled (1), (2), and (3) in \fig{SM::local_TF} show the density profile for a spatially symmetry trap with angular frequency $\bar \omega$ for $T/T_F^\text{trap}=0.2,0.4,$ and $1$, respectively. The colorcoding of the curves indicates the local dimensionless temperature $T/T_F(\mathbf{r})$. It can be seen that within the frame of LDA, for the case of quantum degeneracy in the trap where $T/T_F^{\text{trap}} \lesssim 1$, $T/T_F(\mathbf{r})$ is still larger than 1 for the majority of the density profile.
As demonstrated by the inset of \fig{SM::local_TF} and curve~(2) of \fig{SM::local_TF}, even for $T/T_F^{\text{trap}}=0.4$, the ``hot part" of the cloud extends to $r/R_F$ as small as $\sim0.6$. The inset shows that about half of the cloud is captured by the high-temperature equation of state and the other half by the intermediate-temperature equation of state (purple line). For $T/T_F(\mathbf{r})=0.2$, in contrast, the low-temperature equation of state (blue line) contributes about $30\%$.

\section{Reproduction of the experimental results}
\label{SMsect::experiment}

\begin{figure}[t]
    \centering
    \includegraphics[width=0.45\textwidth]{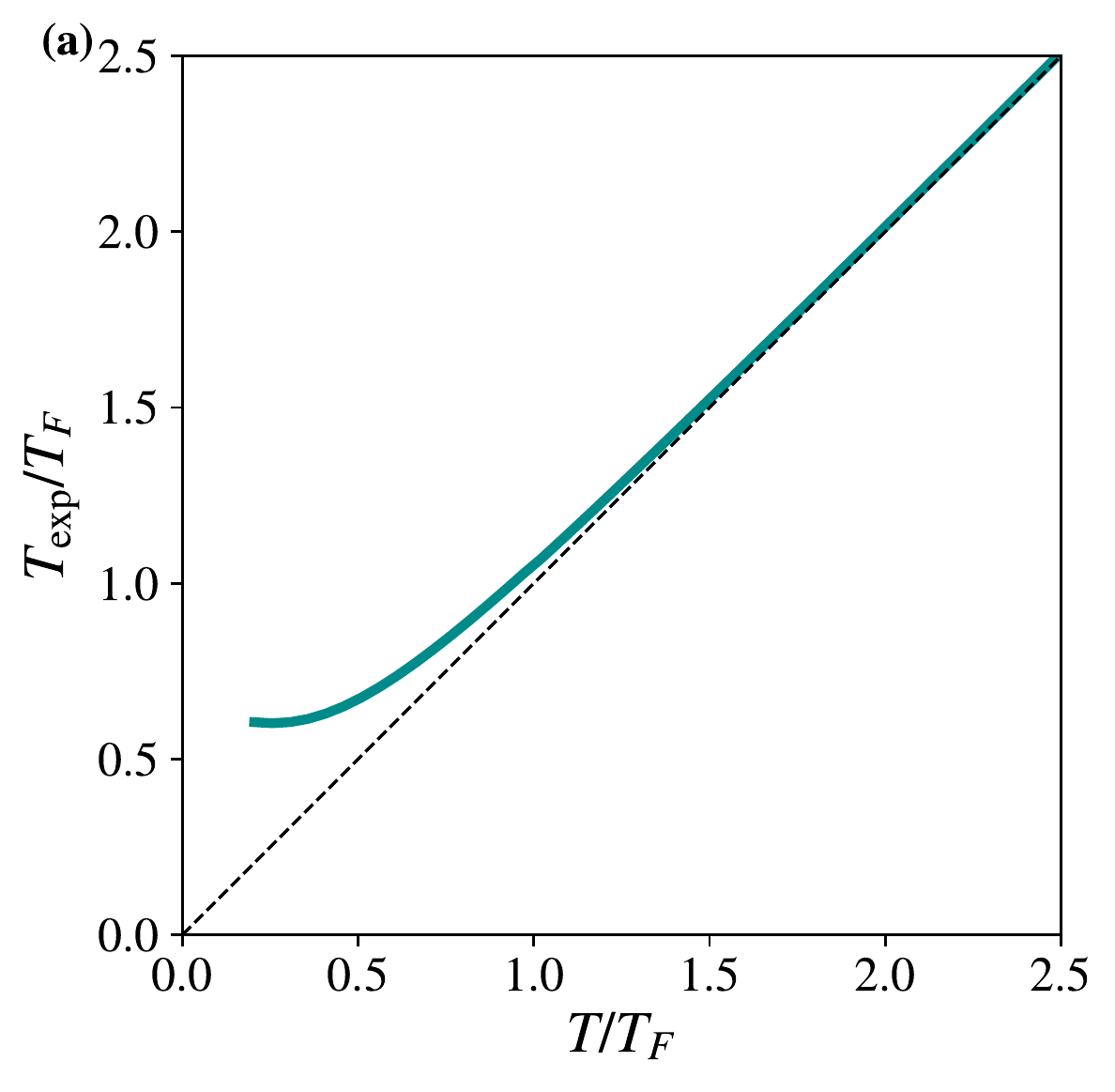}
    \includegraphics[width=0.44\textwidth]{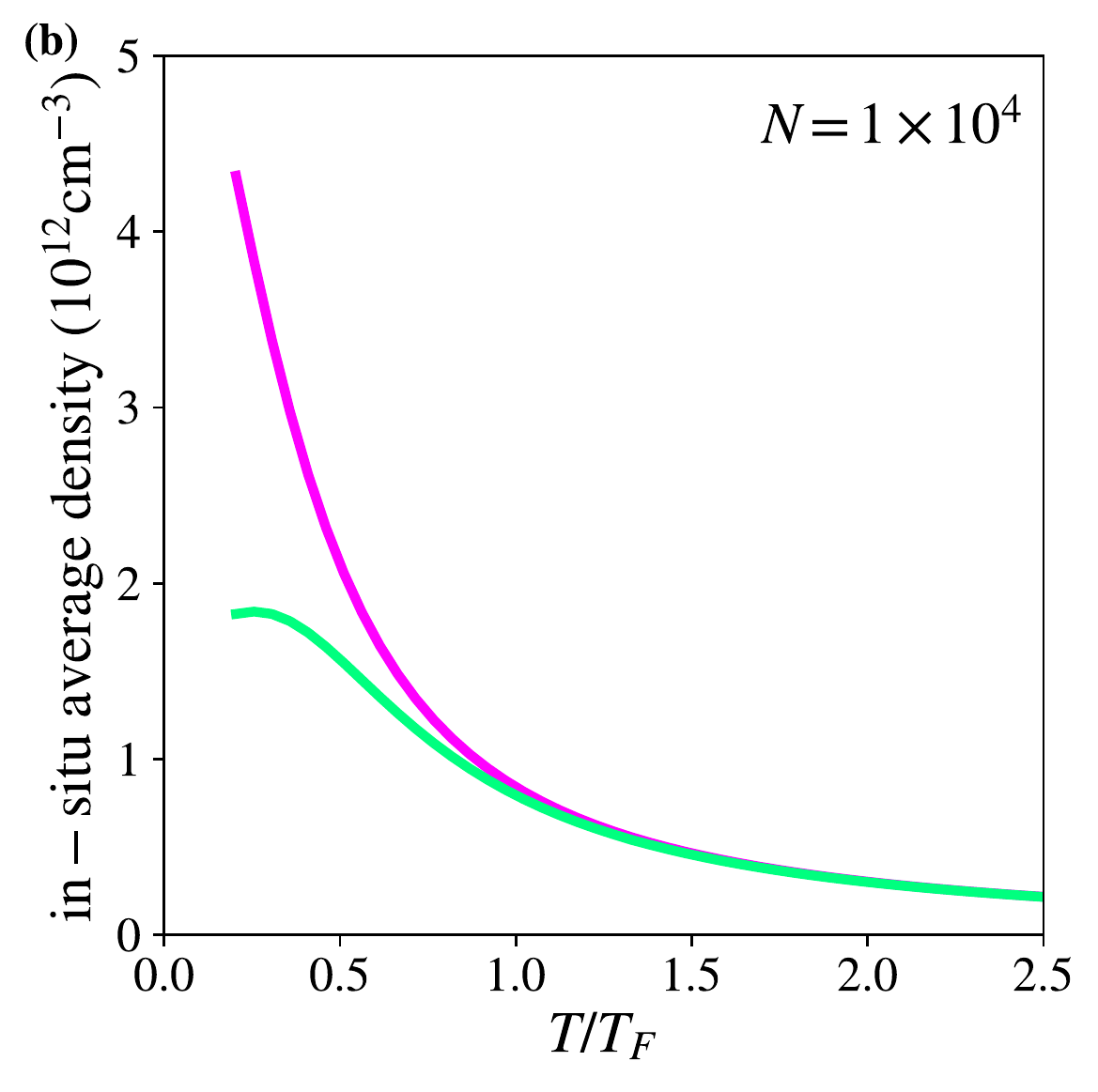}
    \caption{(a) Comparison between "expansion temperature" $T_\mathrm{exp}$ and $T$ scaled by $T_F$. The relation does not depend on the total number of particles $N$ in the system. 
    The dashed black line shows $T_\mathrm{exp}=T$ to guide the eye.
    (b) In-situ average densities against $T/T_F$. Green line represents $\ntrap_\mathrm{exp}$ and magenta line is $\ntrap$. $N$ is set to be $1\times10^4$ for this plot.}
    \label{SM::expcomp} 
\end{figure}

This section establishes an explicit connection between $\beta^{\text{trap}}$ and $\beta_{\text{exp}}$. 
The main text already introduced
that the analysis of the experimental data approximates the in-situ average density by~\cite{demarco2019degenerate}:
\begin{equation}
    \ntrap_\mathrm{exp}(T)=\frac{N}{\VT_\mathrm{exp}}=\frac{N}{8\pi^{3/2}}\omegab^3
    \left(\dfrac{k_B T} {m}\right)^{-3/2},
    \label{SM::nTbarexp}
\end{equation}
where $\omegab$ is the geometric mean of the angular frequencies in the $x$-, $y$-, and $z$-directions. In the following discussion, we replace the experimental trap by an isotropic trap with $\omegab=479$~Hz. 
As can be seen by inserting \eq{SM::nrcls} into $n^\text{trap}$, \eq{SM::nTbarexp} is merely the high-temperature limit of it:
\begin{equation}
    \ntrap=\frac{1}{N}\int\mathrm{d}^3r\left(\dfrac{m^{3/2}\omega^3 N\exp\left(\dfrac{-m\omega^2 r^2}{2k_B T}\right)}{(2\pi k_B T)^{3/2}}\right)^2=\frac{N}{8\pi^{3/2}}\omegab^3\left(\dfrac{k_B T}{m}\right)^{-3/2}.
\end{equation}
Motivated by this, one might speculate that $n_{\text{exp}}^{\text{trap}}(T)$
does not provide a faithful description at lower temperatures and in the degenerate regime. 
However, in the experimental analysis, $n_{\text{exp}}^{\text{trap}}(T)$ is evaluated at the so-called "expansion temperature" $T_\mathrm{exp}$, which was measured by fitting the density profile of the cloud after a long time of flight $\tau$ to the fitting function
\begin{equation}
    n_\mathrm{fit}=n_0\exp\left(\dfrac{m\omegab^2r^2}{2k_B T_\mathrm{exp}(1+(\omegab \tau)^2)}\right),
    \label{SM::nrfit}
\end{equation}
where $n_0$ and $T_\mathrm{exp}$ were treated as fitting parameters. To understand the behavior of $T_{\text{exp}}$, \fig{SM::expcomp}(a) shows $T_\mathrm{exp}/T_F$ as a function of $T/T_F$. 
To obtain $T_\text{exp}$, the experimental sequence is emulated. Specifically, the ballistic expansion is simulated numerically, starting with the confined equilibrated cloud with temperature $T$. The expanded cloud is then fit to \eq{SM::nrfit}. 
For the calculations shown in \fig{SM::expcomp}, $\bar{\omega}\tau=2\pi$ is used. We tested that this value is sufficiently large to obtain converged results.
For $T\gg T_F$, $T_\mathrm{exp}$ obtained from the above approach should be the same as the physical temperature $T$~\cite{pethick2008bose}. 
\Fig{SM::expcomp} shows that this is indeed the case for $T/T_F \gtrsim 1.2$. 
In the quantum degenerate regime, in contrast, $T_\mathrm{exp}$ is higher than the physical temperature $T$ and approaches a constant as $T$ approaches zero.  
Correspondingly, there exists a large deviation between $\ntrap_\mathrm{exp}$ and $\ntrap$ at low temperatures. 
This is demonstrated in \fig{SM::expcomp}(b).

The main text shows that the global loss coefficient $\beta^{\text{trap}}$, which characterizes the trapped system, contains two parts, namely $\beta_{\text{loss}}^{\text{trap}}$ and $\beta_{\text{deform}}^{\text{trap}}$. Correspondingly, the use of $n_{\text{trap}}^{\text{trap}}(T_{\text{exp}})$ also leads to two terms.
Taking the derivative of both sides of \eq{SM::nTbarexp} with respect to $t$, we find 
\begin{equation}
    \frac{\mathrm{d}\ntrap_\mathrm{exp}}{\mathrm{d}t}=-\beta_\mathrm{exp}(\ntrap_\mathrm{exp})^2-\frac{3}{2}\frac{\ntrap_\mathrm{exp}}{T_\mathrm{exp}}\frac{\mathrm{d}T_\mathrm{exp}}{\mathrm{d}t},
    \label{SM::betaexp}
\end{equation}
where we used
\begin{eqnarray}
\beta_{\text{exp}}=
-\frac{V_{\text{exp}}^{\text{trap}}}{N^2} \frac{\mathrm{d}N}{\mathrm{d}t}.
\end{eqnarray}
The second term on the right hand side of \eq{SM::betaexp}
arises from the time dependence of $\VT_\mathrm{exp}$; using $n_{\text{exp}}^{\text{trap}}$, the time dependence is "converted" to a time dependence of $T_\mathrm{exp}$. As the volume of the system expands, $T_\mathrm{exp}$ rises gradually. This was referred to as anti-evaporation in the experimental KRb papers~\cite{ni2010dipolar,demarco2019degenerate}.
While it might be tempting---motivated by the form of the first term on the right hand side of \eq{SM::betaexp}---to identify
$\beta_{\text{exp}}$ with $\beta_{\text{loss}}^{\text{trap}}$, the $\beta_\mathrm{exp}$ extracted from experiment differs from $\beta_\text{loss}^\text{trap}$ since the analysis of the experimental data used approximations for the global density and temperature.  Therefore, in order to use our predictions (namely, $\beta^{\text{trap}}$) to compare with the experiment, we need to find an expression for $\beta_{\text{exp}}$ in terms of $\beta^{\text{trap}}$. This is achieved by solving  \eq{SM::betaexp} for $\beta_{\text{exp}}$,
\begin{eqnarray}
    \beta_\mathrm{exp}=-\frac{1}{(\ntrap_\mathrm{exp})^2}\frac{\mathrm{d}\ntrap_\mathrm{exp}}{\mathrm{d}t}-\frac{3}{2}\frac{1}{\ntrap_\mathrm{exp}T_\mathrm{exp}}\frac{\mathrm{d}T_\mathrm{exp}}{\mathrm{d}t},
    \end{eqnarray}
    and then rewriting terms to bring in $\beta^{\text{trap}}$,
    \begin{align}
    \beta_\mathrm{exp}
    &=-\left[\frac{(\ntrap)^2}{(\ntrap_\mathrm{exp})^2}\frac{\mathrm{d}\ntrap_\mathrm{exp}}{\mathrm{d}\ntrap}\right]\left[\frac{1}{(\ntrap)^2}\frac{\mathrm{d}\ntrap}{\mathrm{d}t}\right]-\frac{3}{2}\left(\frac{T}{T_\mathrm{exp}}\frac{\mathrm{d}T_\mathrm{exp}}{\mathrm{d}T}\right)\left(\frac{1}{\ntrap_\mathrm{exp}T}\frac{\mathrm{d}T}{\mathrm{d}t}\right)\\
    &=\left[\frac{(\ntrap)^2}{(\ntrap_\mathrm{exp})^2}\frac{\mathrm{d}\ntrap_\mathrm{exp}}{\mathrm{d}\ntrap}\right]\betaT-\frac{3}{2}\left(\frac{1}{T_\mathrm{exp}}\frac{\mathrm{d}T_\mathrm{exp}}{\mathrm{d}T}\right)\left(\frac{1}{\ntrap_\mathrm{exp}}\frac{\mathrm{d}T}{\mathrm{d}t}\right). \label{SM::conversion}
\end{align}
\begin{figure}[b]
    \centering
    \includegraphics[width=0.6\textwidth]{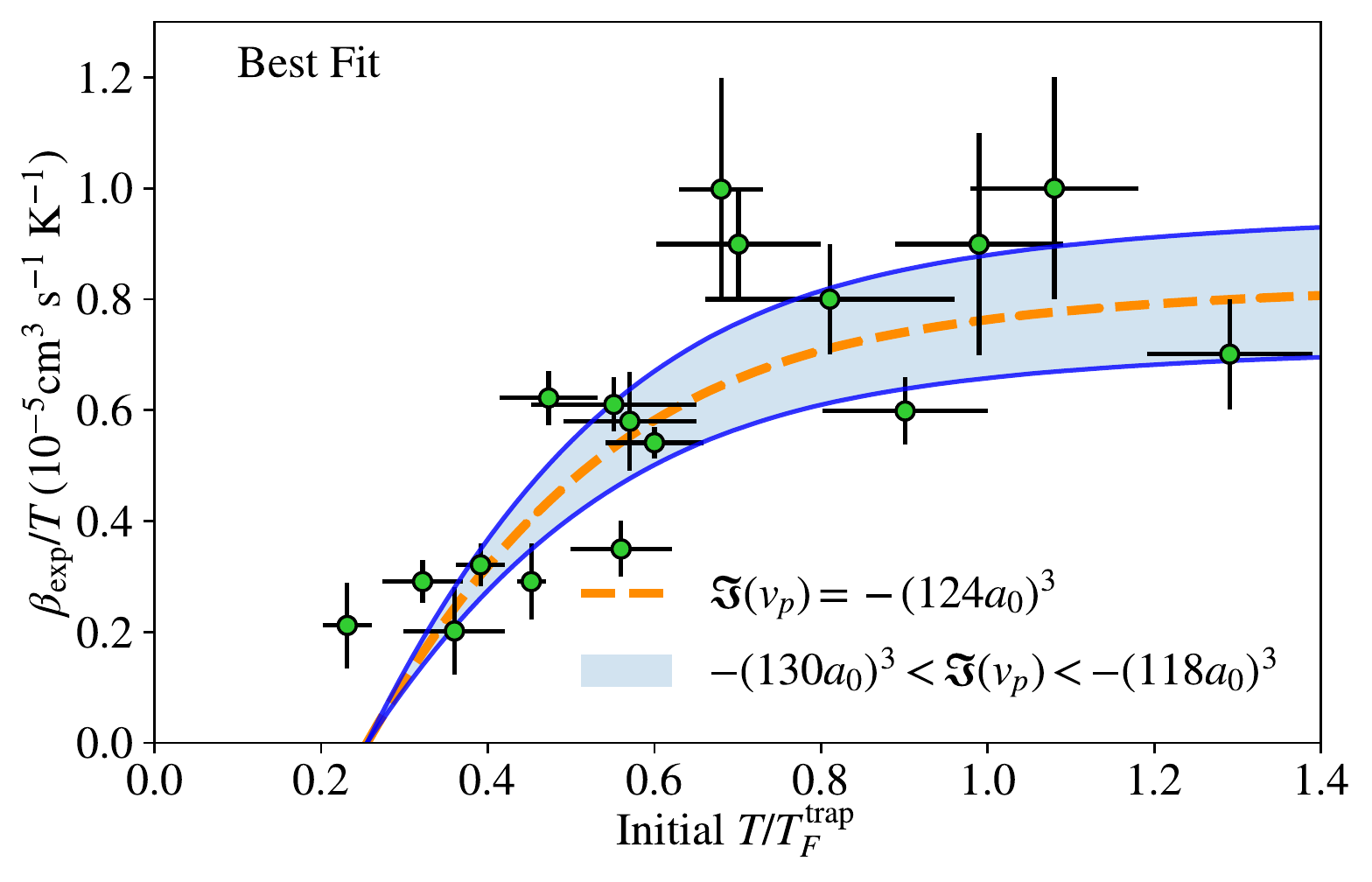}
    \caption{Comparison between the experimental data from Ref.~\cite{demarco2019degenerate} and the theory predictions that use the imaginary part of the scattering volume obtained by fitting the experimental data of Ref.~\cite{demarco2019degenerate}.}
    \label{fig:direct_fitting}
\end{figure}
To reproduce the $\beta_\text{exp}/T$ as a function of the $\textit{initial}~T/T_F$ from Ref.~\cite{demarco2019degenerate}, all $T$, $T_\text{exp}$, and their differentiations are evaluated at $t=0$, i.e., from the equilibrated system before any loss processes have occured.
In fact, this is the only regime in which \eq{SM::conversion} firmly works.
For larger $t$, the density profile may deviate from its shape in thermal equilibrium.
Since cloud deformation effects may be smeared out during the ballistic expansion, in practice one can very likely still obtain $T_\text{exp}$ (and possibly even "$T/T_F$") by fitting the density profile to a polylog function. The extracted $T$, however, might not represent the true temperature due to the fact that temperature is not a well-defined quantity in a non-equilibrated system.
Similarly, we want to emphasize that the quantity ${\mathrm{d}T}/{\mathrm{d}t}$, which enters the calculation, may not correspond to a true physical "heating rate" of the cloud. If the time scale for the system to thermalize is much longer than the period of observation, the quantity ${\mathrm{d}T}/{\mathrm{d}t}$ looses its physical meaning after a certain time.
Hence, it is suggested to interpret it as representing the volume change.
In other words, when treating non-homogeneous systems, where the volume can change, our method serves as a time-dependent perturbation approach.
For our aim, at $t=0$, ${\mathrm{d}T}/{\mathrm{d}t}$ can be evaluated using \eq{SM::betaTV}:
\begin{equation}
    \frac{\mathrm{d}T}{\mathrm{d}t}=\frac{1}{N}\frac{\mathrm{d}\VT}{\mathrm{d}t}\left(\frac{\mathrm{d}\ntrap}{\mathrm{d}T}\right)^{-1}(\ntrap)^2=\beta_\text{deform}^\text{trap}\left(\frac{\mathrm{d}\ntrap}{\mathrm{d}T}\right)^{-1}(\ntrap)^2.
    \label{SM::dTdt}
\end{equation}

According to \fig{SM::expcomp}, $T$ is very close to $T_\mathrm{exp}$ for $T/T_F\gtrsim 1.2$; this tells us that the experimentally measured $\beta_\mathrm{exp}$
coincides with $\beta_\text{loss}^\text{trap}$.
Moreover, because $\beta=\beta_\text{loss}^\text{trap}$ for $T\gg T_F$ (see the main text), the high-$T$ experiments explicitly yield the homogeneous loss coefficient $\beta$.  
Equation~(\ref{highTbeta}) of the main text expresses $\beta$ in terms of the parameter $\Im(v_p)$. An earlier experiment~\cite{ospelkaus2010quantumstate} that operated in the high-temperature regime reported $\beta/T$ to be $1.1(\pm3)\times10^{-5}~\mathrm{cm^3s^{-1}K^{-1}}$. Using the value of $\beta/T$, we extract $\Im(v_p)\simeq-(136^{+11}_{-14}a_0)^3$.

Alternatively and to further verify our results, we perform a non-linear fit to the experimental result of Ref.~\cite{demarco2019degenerate} using Eqs.~(\ref{SM::conversion}) and (\ref{SM::dTdt}). Weighting each experimental data point with $1/[(3\sigma_x)^2+(3\sigma_y)^2]$, we obtain $\Im(v_p)=-(124^{+6}_{-6}a_0)^3$. This best fit value is  consistent with both the predictions from the MQDT calculations and the value extracted from the high-temperature experiment. Figure~(\ref{fig:direct_fitting}) shows the result.

\end{document}